\documentclass[a4paper,12pt]{iopart}
\expandafter\let\csname equation*\endcsname=\relax 
\expandafter\let\csname endequation*\endcsname=\relax 
\usepackage{graphicx,amssymb,color} 
\usepackage[utf8]{inputenc}
\usepackage[english]{babel}

\usepackage{xcolor}

\usepackage{caption}
\usepackage{hyperref}
\usepackage{epsfig}
\usepackage{mathrsfs}
\usepackage{verbatim}
\usepackage{centernot}
\usepackage{siunitx}
\usepackage{physics}
\usepackage{array}
\usepackage{bm}	

\renewcommand{\c}{\hat{c}}

\newcommand{\n}{\hat{n}}
\newcommand{\cd}{\hat{c}^\dagger}

\newcommand{\bd}{\hat{b}^\dagger}
\renewcommand{\b}{\hat{b}}

\newcommand{\U}{\hat{U}}

\renewcommand{\vec}[1]{\bm{#1}}

\newcommand{\detuning}{\Delta}

\begin{document}

\title{Self-generated quantum gauge fields in arrays of Rydberg atoms}

\author{Simon Ohler$^1$, Maximilian Kiefer-Emmanouilidis$^{1,2}$, Antoine Browaeys$^3$, Hans Peter B\"uchler$^4$, and Michael Fleischhauer$^1$ }
\address{$^1$Department of Physics and Research Center OPTIMAS, University of Kaiserslautern, 67663 Kaiserslautern, Germany\\
$^2$Department of Physics and Astronomy, University of Manitoba, Winnipeg R3T 2N2, Canada\\
$^3$Universit\'e Paris-Saclay, Institut d’Optique Graduate School, CNRS, Laboratoire Charles Fabry,
91127 Palaiseau Cedex, France\\
$^4$Institute for Theoretical Physics III and Center for Integrated Quantum Science and Technology,
University of Stuttgart, 70550 Stuttgart, Germany}

\begin{abstract}
As shown in recent experiments [V. Lienhard \textit{et al.}, Phys.~Rev.~X~\textbf{10}, 021031 (2020)],  
spin-orbit coupling in systems of Rydberg atoms can give rise to density-dependent Peierls Phases in second-order hoppings  of Rydberg spin excitations and nearest-neighbor (NN) repulsion.  We here study theoretically a one-dimensional zig-zag ladder system of such 
spin-orbit coupled Rydberg atoms at half filling.
The second-order hopping is shown to be associated with an effective gauge field, which in mean-field approximation 
is static and homogeneous.
Beyond the mean-field level the
gauge potential attains a transverse quantum component whose amplitude is dynamical and linked to density modulations. 
We here study the effects of this to the possible ground-state phases of the system.
In a phase where strong repulsion leads to a density wave, we find that as a consequence of the induced quantum gauge field a regular pattern of current vortices is formed. However also  in the absence of density-density interactions the quantum gauge field attains a non-vanishing amplitude. Above a certain critical strength of the second-order hopping 
the energy gain due to gauge-field induced transport overcomes the energy cost from the associated build-up of density modulations leading to a spontaneous generation of the quantum gauge field.
\end{abstract}

\pacs{123}

\date{\today}
\maketitle

\section{Introduction}

Due to their strong and non-local interaction, their experimental accessibility and the high degree of tunability and control that can be exerted, Rydberg atoms have become a powerful tool for simulating strongly correlated quantum many-body systems \cite{Browaeys2020}.  The strong interaction exhibited by Rydberg atoms extending over $\mu$m distances, while at the same time featuring high stability, allows to use optical tweezers and atom-by-atom assembly to prepare  arbitrary one- and two-dimensional arrays of atoms \cite{Endres2016,Barredo2016,Barredo2018}. The repulsive van-der-Waals interaction between pairs of Rydberg atoms then provides a natural way to implement spin models with Ising-type $zz$ interactions \cite{Labuhn2016} and to study them with full coherent control. Experiments include the observation of quantum phase transitions to ordered phases in one-dimensional quantum models with Ising interactions extending beyond the nearest neighbor \cite{Schauss2015,Zeiher2017,Bernien2017} and two-dimensional square lattices \cite{Ebadi2020,Scholl2021,GuardadoSanchez2018,Lienhard2018}, as well as the probing of topological spin liquids in dimer models based on Rydberg atoms in Kagome lattices \cite{Semeghini2021}. 
Here the spin degree of freedom is formed by a low-lying level and a Rydberg state of the atoms. If the spin is made up out of two Rydberg states with a dipole-allowed transition, excitation exchange between pairs of atoms corresponds to an $xy$ spin interaction, which has been used to experimentally investigate symmetry protected topological lattice models of interacting bosons \cite{Leseleuc2019}.  

The realization of external magnetic fields pose a challenge to all platforms based on neutral atoms. To circumvent this problem, different physical effects that mimic the behaviour of a magnetic field on charged particles have been proposed including laser assisted tunneling \cite{Jaksch2003}, periodic driving \cite{GoldmanPRX2014} or the transfer of angular momentum from light \cite{Goldman2014}. Recently, in \cite{Lienhard2020} it was demonstrated that spin-orbit coupling in systems of Rydberg atoms can be used to generate Peierls phases in the hopping matrix elements of Rydberg excitations
without external light fields, laser assisted tunneling or lattice
shaking. The authors showed that the Peierls phases, which are the manifestation of external magnetic fields in lattice models, 
lead to chiral currents in a triangle-configuration of three atoms. Moreover the complex amplitudes of the hopping process between two Rydberg atoms depended on the occupation of Rydberg states of a third atom in between the two atoms. Such density-controlled hopping processes form one elementary building block for a direct implementation of even more complex many-body models such as dynamical gauge theories \cite{Wiese2013,Dalmonte2016}. 

In the present paper we theoretically investigate a simple one-dimensional extension of the triangle system studied in \cite{Lienhard2020} with density-dependent complex hopping of Rydberg excitations  competing with repulsive density-density interactions. We will show that this model features besides a static, classical gauge potential a dynamical quantum component linked to density modulations. Although being a dynamical quantity, the quantum component of the gauge field is  gauge 
invariant as it is a transverse field. Nevertheless, it substantially modifies the ground-state phase diagram, which we numerically investigate using exact diagonalization methods. It leads to two new liquid phases in addition to a
trivial superfluid, which are characterized by alternating vortex currents of Rydberg excitations. In an ordered phase where the
repulsive interactions dominate a regular array of current vortices emerges. Surprisingly the quantum gauge field is spontaneously generated even without additional density-density interaction terms, reminiscent of the Higgs-mechanism. 

The paper is organized as follows:
In Section \ref{section:microphysics} we shortly illustrate the microscopic origin of the model. In the subsequent Section \ref{section:SpinExcitations_1D} we discuss the Rydberg spin model and show its phase diagram as well as the emergence of chiral currents. In order to gain some understanding of the phase diagram we transform the spin model, equivalent to hard core bosons,
to a fermion model and discuss a mean-field approximation of the fermionic model in 
Section \ref{section:mean_field_hamiltonian}. In Section \ref{section:emerging_gauge} we reformulate the model in terms of operator valued
transverse gauge fields which provides an explanation for the vortex-currents found in the numerical simulations. In Section \ref{sec:exp_realization} we make a few remarks about possible experimental signatures and in Section \ref{sec:summary} we summarize our results and give an outlook to further studies.

\section{Nonlinear excitation transport in arrays of Rydberg atoms}\label{section:microphysics}
\begin{figure}[htb]
	\begin{center}
	\includegraphics[width=0.7\columnwidth]{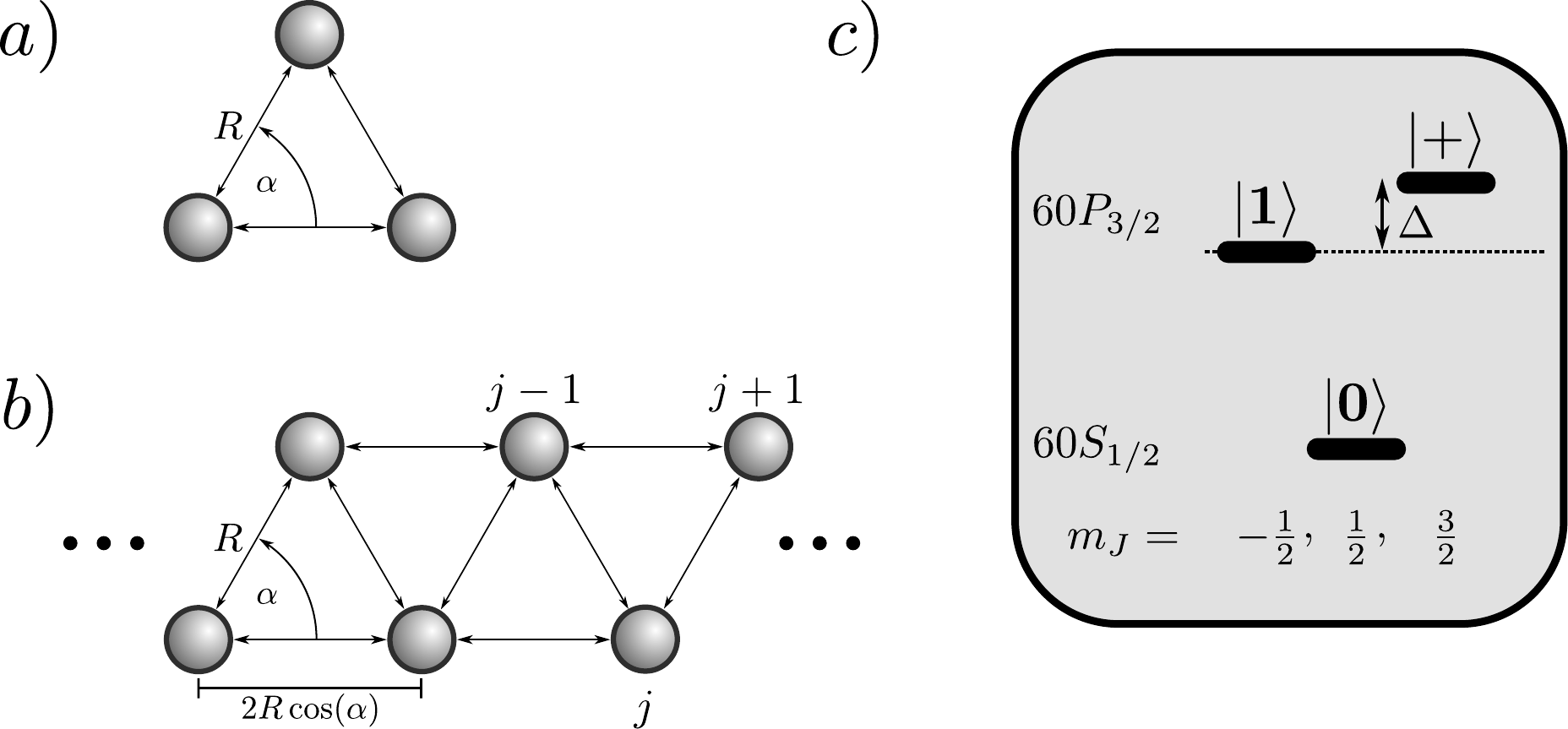}
	\end{center}
	\caption{a): Minimal setup of Rydberg atoms in the $(x,y)$-plane subjected to a magnetic field in $z$-direction. b): Zig-zag chain discussed in this paper. The distance R and angle $\alpha$ between nearest neighbors determine the strength and complex phase of the interaction processes. c): Relevant atomic levels, which we take from a Rydberg S $(\ket{0})$ and P $(\ket{+},\ket{1})$ orbital. In the experiments of \cite{Lienhard2020} the states $60S_{1/2}$ and $60P_{3/2}$ are used.}
	\label{fig:same_species_all}
\end{figure}
%
To understand how Rydberg interactions can be tuned to exhibit density-dependent Peierls phases, we first consider the physical behaviour of a minimal triangle setup of atoms, shown in Fig.~\ref{fig:same_species_all}. In this setup, three identical Rydberg atoms are placed on an equilateral triangle in the xy-plane. Furthermore, two Rydberg states are chosen to represent the two spin-states $\ket{1}$ and $\ket{0}$. The third atomic state, $\ket{+}$, is detuned and will be adiabatically eliminated, yielding non-resonant processes that scale as the inverse detuning $1/\Delta$. The  level scheme shown in Fig.~\ref{fig:same_species_all} is created by external electric and magnetic fields which provide the required energy splittings.
\\
When the magnetic field is orthogonal to the plane of the atoms, the microscopic Hamiltonian coupling atoms $i$ and $j$ can be written as
\begin{align}\label{eqn:V_ij_finalform}
\hat{V}_{ij}&=\frac{1}{4\pi\epsilon_{0}R_{ij}^{3}}\left[
\hat{d}^{z}_{i}\hat{d}^{z}_{j}+\frac{1}{2}\left(\hat{d}^{+}_{i}\hat{d}^{-}_{j}+\hat{d}^{-}_{i}\hat{d}^{+}_{j}\right)-\frac{3}{2}
\left(
\hat{d}^{+}_{i}\hat{d}^{+}_{j}\mathrm{e}^{-2i\phi_{ij}}+\hat{d}^{-}_{i}\hat{d}^{-}_{j}\mathrm{e}^{2i\phi_{ij}}
\right)
\right].
\end{align}
Here, $R_{i,j}=\abs{\vec{r}_{i,j}}$ and $\phi_{ij}$ refer to the interatomic distance and angle in polar coordinates, respectively. Additionally, the operators $\hat{d}^{\pm}_{i}=\mp\frac{1}{\sqrt{2}}(\hat{d}^{x}_{i}\pm i\hat{d}^{y}_{i})$ represent ladder operators raising and lowering the angular momentum of the $i$-th Rydberg atom. In \eqref{eqn:V_ij_finalform} we recognize the flip-flop terms $\hat{d}^{+}_{i}\hat{d}^{-}_{j}$ as well as terms $\hat{d}^{+}_{i}\hat{d}^{+}_{j}e^{-2i\phi_{ij}}$ that increase the angular momentum of both involved atoms and are multiplied by a complex phase.
\\
As a result of the interaction Hamiltonian \eqref{eqn:V_ij_finalform} it is useful to consider two different types of processes. In this section, we will only provide a qualitative explanation, the microscopic derivation is given in \ref{sect:App_MicroDerivation}.
\\
We consider the states $\ket{1}$ and $\ket{0}$ as the two states of a spin-1/2 system and transform to hard-core bosons  for a more intuitive picture.
%
\begin{figure}[htb]
	\begin{center}
	\includegraphics[width=0.7\columnwidth]{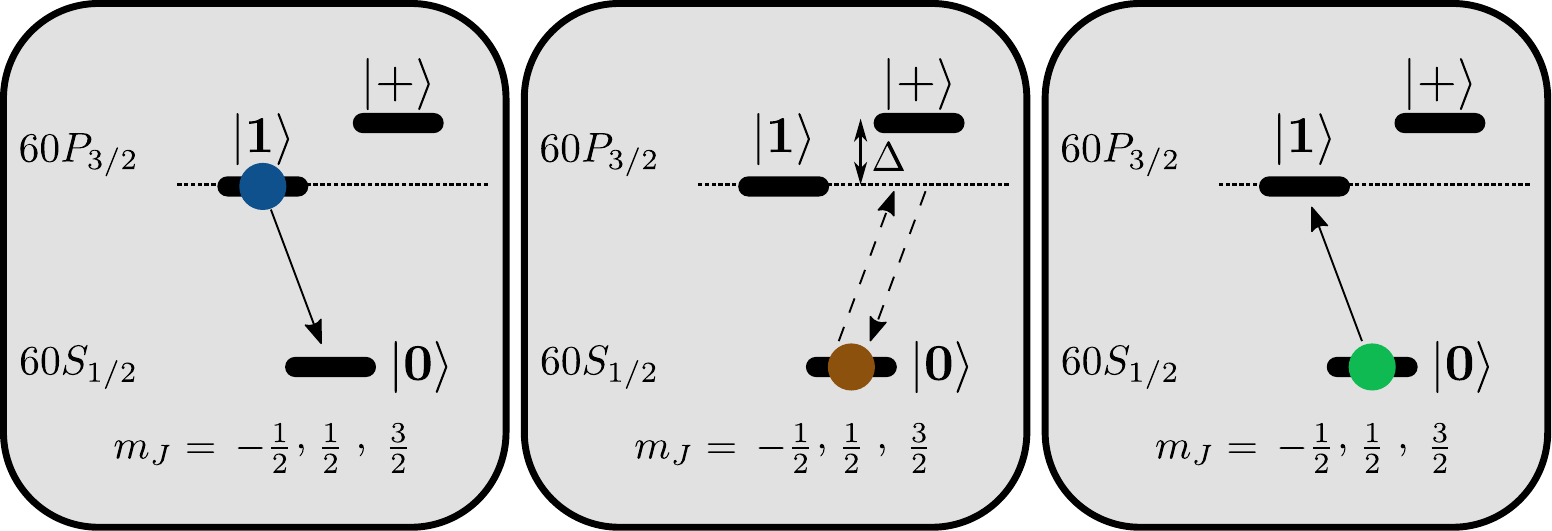}
	\end{center}
	\caption{Illustration of the indirect process occurring in the triangle of Rydberg atoms. The three boxes correspond to the operation in each of the three atoms. Atom 1 (blue) is de-excited, where the photon couples to the $\ket{0}\leftrightarrow \ket{+}$ transition in atom 2 (brown). Since this transition is detuned, the process is non-resonant and returns to the state $\ket{0}$, which in turn couples to the resonant $\ket{0}\leftrightarrow \ket{1}$ transition of the third (green) atom.}
	\label{fig:IndirectProcess_visual}
\end{figure}
%
The first and simplest process is the resonant dipole-dipole interaction, where the term $\hat{d}^{+}_{i}\hat{d}^{-}_{j}$ in \eqref{eqn:V_ij_finalform} causes an exchange of population in the $\ket{0}$ and $\ket{1}$ states of the atoms $i$ and $j$. This mechanism yields for the hard-core bosonic Hamiltonian
 $\hat{d}^{+}_{i}\hat{d}^{-}_{j} \rightarrow \hat{b}^{\dagger}_{j}\hat{b}_{i}$,
where the operators $\hat{b}_{i}$ and $\hat{b}^{\dagger}_{i}$ are annihilation and creation operators for hard-core bosons (spin-1/2) at site $i$, respectively, which satisfy the hard-core constraint $\hat{b}_{i}^{2}=\bigl(\hat{b}^{\dagger}_{i}\bigr)^{2}=0$ at each site $i$.
\\
The second, more interesting process is the non-resonant excitation transfer involving three sites of the lattice. For this process, illustrated in Fig.~\ref{fig:IndirectProcess_visual}, the terms quadratic in raising or lowering operators in \eqref{eqn:V_ij_finalform} are responsible. An important feature is here that de-excitation of the blue atom drives the non-resonant process in the red atom only if the latter is in the ground state $\ket{0}$.  In this manner, the hopping process shown in Fig.~\ref{fig:IndirectProcess_visual} connects two sites (blue and green atoms) depending on the density (occupation of state $\ket{1}$) of another atom (red). Since the corresponding terms of \eqref{eqn:V_ij_finalform} also include a complex phase,  we obtain the following term in hard-core boson language
    $\hat{b}^{\dagger}_{j}\hat{b}_{i}\left(1-\hat{n}_{k}\right)\mathrm{e}^{-4 i\epsilon_{i,j,k}\alpha}$.
Here, $\epsilon_{i,j,k}$ is the Levi-Civita symbol which distinguishes between a clockwise and counter-clockwise motion of the excitation, and $\alpha=\phi_{ij}$ is the interatomic angle for all $i\neq j$.
\\
Lastly, we need to take into account a special case of the second process, where the initial and the final atoms are identical. In that situation, we no longer have a hopping process, but in fact a nearest-neighbor density-density interaction
 $   \hat{b}^{\dagger}_{i}\hat{b}_{i}\left(1-\hat{n}_{k}\right)$,
where the complex phase drops out.
\\
Summarising all the terms discussed so far, we can write down the total Hamiltonian for the triangle setup discussed in Fig.~\ref{fig:same_species_all}
\begin{align}\label{eqn:basic_triangle_hamiltonian}
\hat{H}&=-J\sum_{i\ne j=1}^{3}\bd_{i}\b_{j}+h.c.
-2gJ\sum_{i\neq j\neq k}^{3}
\bd_{i}\b_{j}\left(1-\n_{k}\right)\mathrm{e}^{-4 i\epsilon_{i,j,k}\alpha}+h.c.
-
2gJ\sum_{i\neq k}^{3}\n_{i}\left(1-\n_{k}\right).
\end{align}
Here, $J$ is the natural energy scale of the system and $g=\frac{27J}{2\Delta}$ is the rescaled detuning of the atomic state $\ket{+}$. The numerical factors in $g$ are explained in the microscopic derivation given in \ref{sect:App_MicroDerivation}. This model has been experimentally investigated by the authors of \cite{Lienhard2020}, where they confirmed the density-dependence of the chiral motion of excitations on the triangular setup.
\\
In the following we consider an extension of this model that extends in 1D in the form of a zig-zag chain, as shown in Fig.~\ref{fig:same_species_all}. As a result, more density-dependent terms and longer-range processes enter the Hamiltonian given in \eqref{eqn:basic_triangle_hamiltonian}, but the underlying principle remains unchanged.

\section{Spin excitations in a 1D zig-zag chain}\label{section:SpinExcitations_1D}

\subsection{Model}

\begin{figure}[ht]
	\includegraphics[scale=0.5]{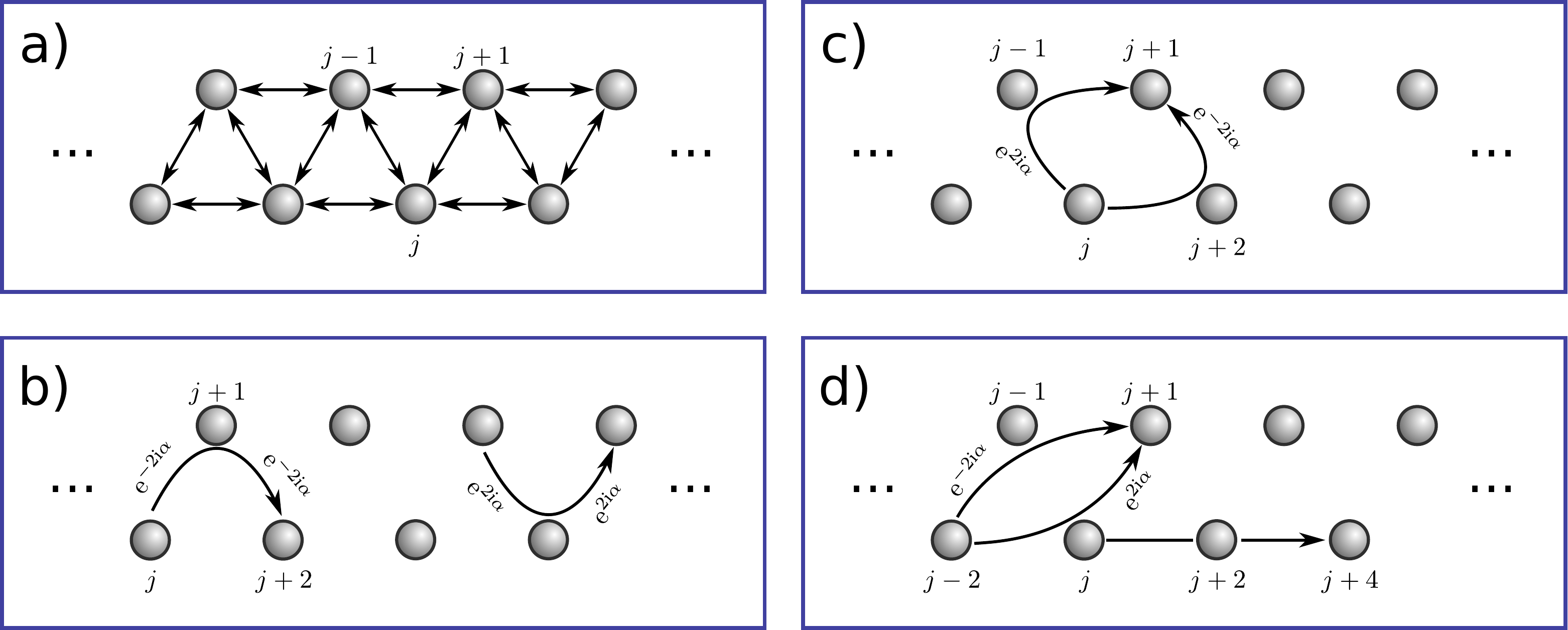}
	\caption{Visualization of all hopping processes in the one-dimensional zig-zag ladder, including their respective 
	phases. (a) Direct hopping processes. (b) Indirect hoppings between NN on the same sub-chain, acquiring opposite Peierls phases depending on the direction. (c) Two exemplary indirect NN hoppings connecting sub-chains. The total Peierls phase attached to such a process is staggered, i.e. it changes sign when going from the upper to the lower chain or vice versa. The same holds for all other non-resonant processes. (d) Indirect hopping terms connecting sites $j-2$ with $j+1$ as well as $j$ with $j+4$. For the case of $\alpha=\frac{\pi}{3}$ chosen here, all these indirect terms are of equal magnitude.}
	\label{fig:1D_Model_allTerms}
\end{figure}

We now consider a one-dimensional zig-zag chain formed by the elementary triangles of Rydberg atoms from the previous section, see Fig.~\ref{fig:same_species_all}b). We set the physical angle $\alpha=\pi/3$ in which case the triangles are equilateral, and each site $j$ now has four equivalent nearest neighbors (NN). As a result, the indirect (density-dependent) hopping terms as well as the NN interaction outlined in the previous section now include all possible NN-combinations as shown in Fig.~\ref{fig:1D_Model_allTerms}. 
In order to account for the fact that more internal Rydberg states may be relevant in the interaction we 
allow for additional NN density-density interactions resulting from off-resonant van-der Waals interactions. Their relevance can in fact be controlled by proper choice of Rydberg states. We here quantify them by a dimensionless parameter $\eta$, where $\eta=1$ corresponds to the case when no additional van-der Waals interactions are present (i.e. the case of a three-level atom), and $\eta \gg 1$ means that van-der Waals interactions dominate over the second order processes discussed in the previous section. Likewise it is possible to tune the
levels to a sweet spot where van-der Waals level shifts cancel, i.e. $\eta=0$. Note that $\eta$ depends on the principal quantum number of the Rydberg level and thus is not a continuously tuneable parameter.
The complete many-body Hamiltonian for the zig-zag chain then reads
\begin{eqnarray}
\hat{H}&=&-J\sum_{j}\bd_{j+1}\b_{j}
\left[1 + 2 g 
\left(e^{\mp \frac{2\pi i}{3}} (1-\n_{j-1})
+e^{\pm \frac{2\pi i}{3}} (1-\n_{j+2}) 
\right)
\right]+h.c.
\nonumber\\
&&-J\sum_{j}\bd_{j+2}\b_{j}
\left[1+2 g\, e^{\pm \frac{4\pi i}{3}}
(1-\n_{j+1})
\right]+h.c.
\label{eq:H}\\
&& -J\sum_{j}\bd_{j+3}\b_{j} \,  2 g
\left[e^{\mp \frac{2\pi i}{3}}(1-\n_{j+1}) +
e^{\pm \frac{2\pi i}{3}}(1-\n_{j+2})
\right]+h.c.
\nonumber \\
&& -J\sum_{j}\bd_{j+4}\b_{j} \,  2 g
(1-\n_{j+2})+h.c.
\nonumber\\
&&-J\sum_{j}\bd_{j}\b_{j}\,  2 g\eta \left[
(1-\hat{n}_{j-1})+(1-\hat{n}_{j+1})+(1-\hat{n}_{j-2})+(1-\hat{n}_{j+2})
\right].\nonumber
\end{eqnarray}
The upper (lower) sign in the exponents applies to even (odd) sites $j$ corresponding to the upper (lower) sub-chain. $\hat b_j, \hat b^\dagger_j$ are annihilation and creation operators of hard-core bosons and $\hat n_j=\bd_j \b_j$ is the number operator of particles acting at site $j$.
The individual hopping processes are illustrated in Fig.~\ref{fig:1D_Model_allTerms}.
$g= 27 J/\left(2\detuning\right)$ is the coupling constant of the second order processes relative to the direct hopping.
Its magnitude determines the relevance of both the complex, density-dependent hopping processes and the density-density interaction as compared to the direct hopping processes occurring at rate $J$.
%
%

The Hamiltonian \eqref{eq:H} conserves the number of hard-core bosons (Rydberg excitations in state $\ket{1}$), and of particular interest is the case of an excitation number commensurate with the lattice size $L$. Consequently, in the following we consider half filling, i.e. the total number of atoms in Rydberg states $\ket{0}$ and $\ket{1}$ is each one half the number of lattice sites.

\subsection{Phase diagram}
\label{sec:phase-diagram}

In the present paper we are interested in the properties of the many-body ground state of the zig-zag chain \eqref{eq:H} for a half-filled system of hard-core bosons with repulsive interactions, i.e. $g>0$ and $\eta\ge 0$. We start by considering the two trivial limiting cases.

\begin{figure}[htb]
	\begin{center}
	\includegraphics[width=\columnwidth]{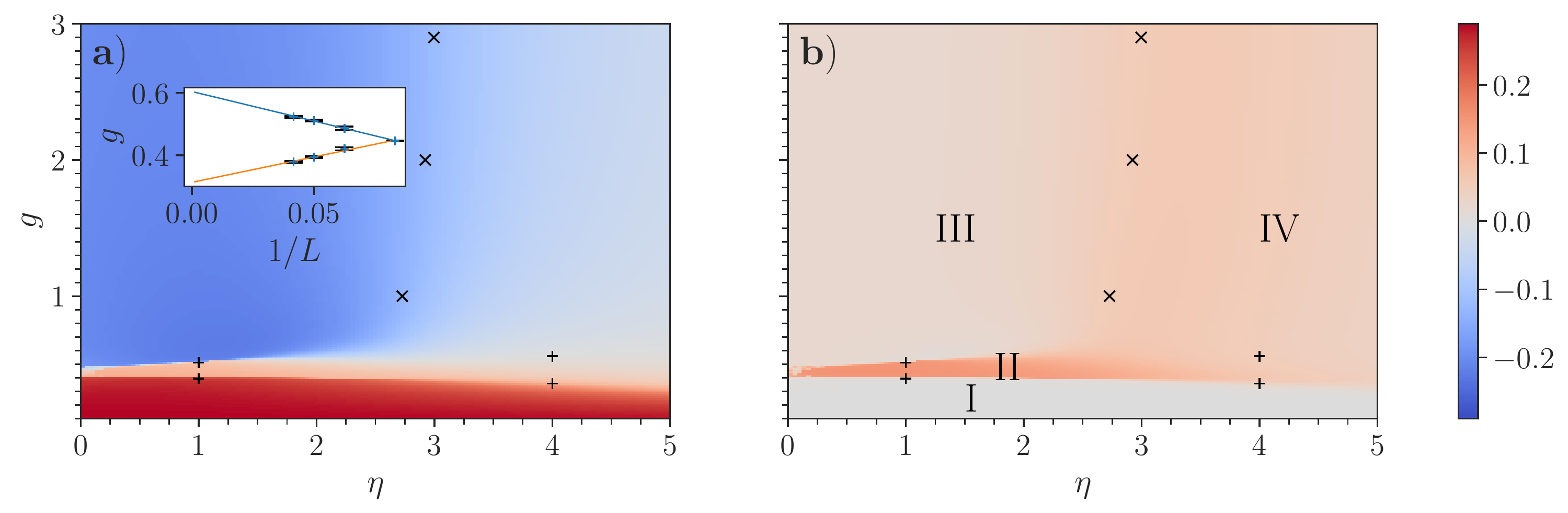}
	\end{center}
	\caption{Phase diagram as a function of $g$ and $\eta$. The  phases I-IV are characterized by a different behavior of single particle correlations. (a) Re$[\bd_{j+1} \b_j]$, (b) Im$[\bd_{j+2} \b_j]$.  
	The cross (plus) signs show the positions of the peaks of the ground-state fidelity obtained from ED simulation in horizontal (vertical) cuts. The inset in (a) shows a finite-size extrapolation of the peak positions for $\eta=1$, demonstrating that the liquid regime II, absent in a mean-field model, survives in the thermodynamic limit $L\to\infty$.
	}
	\label{fig:phase-diagramm}
\end{figure}

If all second-order processes are weak, i.e. for $g\to 0$, which is the case for 
large detuning $\vert \detuning\vert \gg J$, 
only direct hopping processes between nearest neighbors $j\to j+1$ and $j\to j+2$ are relevant and one expects a simple bosonic (quasi) superfluid (phase I in Fig.~\ref{fig:phase-diagramm}).

Controlled by the parameter $\eta$, there is a competition between the density-density interaction, preferring an ordered state for $\eta g\gg 1$  (phase IV in Fig.~\ref{fig:phase-diagramm}), and hopping processes which drive the system into a liquid state (phases I - III). For small values of $\eta$ there is furthermore a competition between direct hopping processes and second-order hopping processes with density-dependent Peierls phases, controlled by $g$.  Of particular interest
are effects caused by the complex, density-dependent hopping terms.

To attain a qualitative overview of the transition points between these phases
we have calculated the dimensionless ground-state fidelity \cite{Zanardi2006}
\begin{equation}
    f(\lambda)=\frac{2}{N} \frac{1-\abs{\bra{\Phi_0(\lambda)}\ket{\Phi_0(\lambda+\delta \lambda)}}}{\delta\lambda^2},\qquad \delta\lambda\to 0,
    \label{eqn:fidelity_definition}
\end{equation}
where $\vert\Phi_0(\lambda)\rangle$ is the many-body ground state of the system with $N$ particles, and $\lambda$ denotes a system parameter ($g$ or $\eta$) that is varied, while keeping the other constant. 
In the thermodynamic limit $L\to\infty$ while keeping $L/N=2$ and within the same phase $f$ is a smooth function of the system parameter(s) $\lambda$ but develops a singularity at a phase transition. In a finite system the ground-state fidelity remains finite but shows a peak, which can still be used as a good indicator of a phase transition.
\\
Fig.~\ref{fig:phase-diagramm} shows a qualitative phase diagram as a function of $g$ and $\eta$. Shown are different first-order correlations in a false colour plot. The crosses/plusses correspond to peaks of the ground-state fidelity $f$ obtained from exact diagonalization (ED) of Hamiltonian \eqref{eq:H} in a finite system of length $L=20$ with periodic boundary conditions, whereas the inset shows the result of a finite size expansion. The latter demonstrates that the intermediate, liquid regime II survives in the thermodynamic limit.

In order to characterize the different phases we discuss in the following density correlations and currents.

\subsection{Density correlations}
\label{sec:density-correlations}

As can be seen from Fig.~\ref{fig:phase-diagramm} the critical value of 
$\eta$ needed for a transition into a phase with density order is larger than unity and on the order of $3$. Thus further density-density interactions are needed in addition to the one associated with the second order hopping processes discussed in Sec.~\ref{section:microphysics}.

\begin{figure}[htb]
	\begin{center}
		\includegraphics[width=\columnwidth]{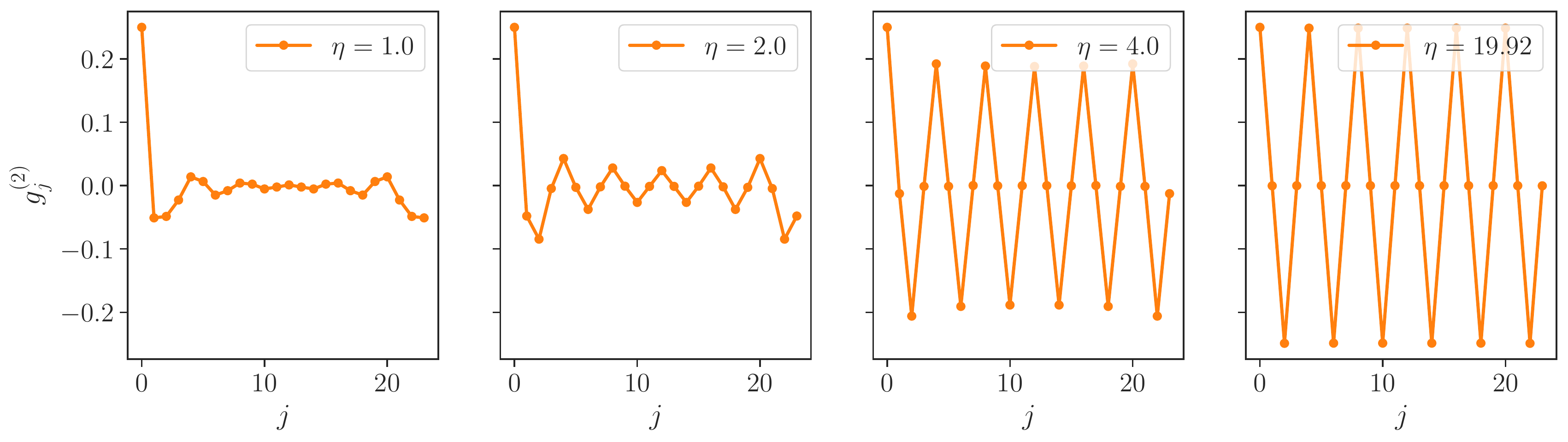}
	\end{center}
	\caption{Density-density correlations for $g=2$ and varying $\eta$ for periodic boundary conditions (PBC), using a system size of $L=24$ sites. One recognizes a transition from a liquid to an ordered phase with half filling of each the upper and lower sub-chain for increasing values of $\eta$. Note that two degenerate states (e.g. $\vert \psi_0^{(1)}\rangle$ and $\vert \psi_0^{(4)}\rangle$) contribute to $g^{(2)}$, explaining the pattern seen for large values of $\eta$.}
	\label{fig:density-correlations}
\end{figure}

In Fig.~\ref{fig:density-correlations} we plotted 
the density-density correlation $g^{(2)}_j =\langle \hat n_0 \hat n_j\rangle -\langle \hat n_0\rangle\langle \hat n_j\rangle$ for different values of $\eta$ at $g=2$.  One recognizes the emergence of long-range density order in both sub-chains when $\eta >  3$. In an infinite system or for periodic boundary conditions and a number of lattice sites that is a multiple of 4, the ordered ground state is 4-fold degenerate:
\begin{eqnarray}
\vert \psi_0^{(1)}\rangle &=&\vert 1\rangle\vert 1\rangle\vert 0\rangle\vert 0\rangle
\vert 1\rangle\vert 1\rangle\dots,\nonumber\\
\vert \psi_0^{(2)}\rangle &=&  \vert 0\rangle\vert 1\rangle\vert 1\rangle\vert 0\rangle\vert 0\rangle\vert 1\rangle\dots,\label{eq:states}\\
\vert \psi_0^{(3)}\rangle &=& \vert 0\rangle\vert 0\rangle\vert 1\rangle\vert 1\rangle\vert 0\rangle\vert 0\rangle\dots,\nonumber\\
\vert \psi_0^{(4)}\rangle &=& \vert 1\rangle\vert 0\rangle\vert 0\rangle\vert 1\rangle
\vert 1\rangle\vert 0\rangle\dots.
\nonumber
\end{eqnarray}
In these configurations every excited site has exactly one excited neighbor, minimizing the energy from density-density interactions. In any other configuration the number of excited neighbors is increased, leading to a larger interaction energy per particle. Thus $g_j^{(2)}$ oscillates with a period of 4. 

\subsection{Liquid phases and chiral average currents}

\begin{figure}[htb]
	\begin{center}
	\includegraphics[width=\columnwidth]{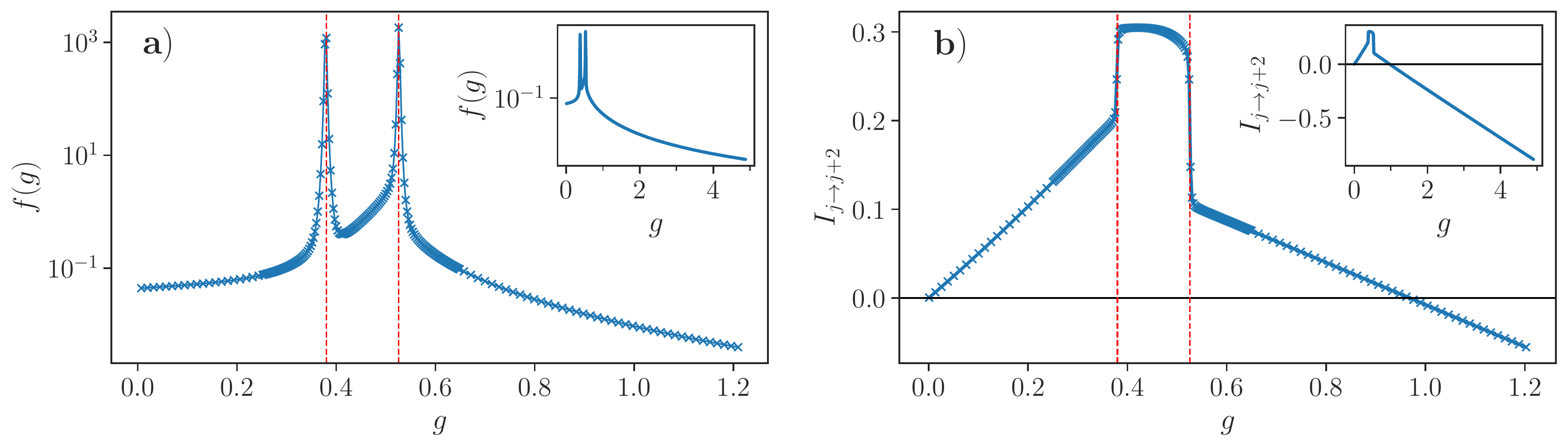}
	\end{center}
	\caption{(a) Ground-state fidelity $f$ (eqn. \eqref{eqn:fidelity_definition}) along a vertical cut through Fig.~\ref{fig:phase-diagramm} at $\eta=1$ and $L=24$ for PBC. One recognizes two peaks, which are indicative of a phase transition. For these peaks, finite-size expansion yields the critical values of $g=0.605$ and $g=0.313$. The inset demonstrates that further increasing $g$ does not change the properties of the system. (b) Currents $I_{j\to j+2}$ for $j=0$ (sub-chain A) as a function of $g$ within the liquid phase for $\eta=1$. The values of $g$ at which $I_{j\to j+2}$ is discontinuous are identical to the peaks in (a).
	}
	\label{fig:GS-fidelity}
\end{figure}

More interesting than the liquid-insulator transition from phase III into phase IV seen in  Fig.~\ref{fig:density-correlations} are those within the liquid phases. Fig.~\ref{fig:GS-fidelity} shows a plot of $f$ along a vertical cut of Fig.~\ref{fig:phase-diagramm} for $\eta=1$. One recognizes two further peaks, one at $g\approx 0.5$ and another one at $g\approx 0.38$. For $g\to 0$ (phase I) the system is in a simple superfluid state, while the other two phases II and III are characterized by non-vanishing currents.

%
Since the complex hopping amplitudes discussed in Sec.\ref{section:microphysics} act analogously to a magnetic field, it is instructive to consider particle currents in the plane of the zig-zag chain. 
To simplify the discussion let us first replace the number operators in eq.\eqref{eq:H} by expectation
values. In this mean-field description there is a non-vanishing Peierls phase for hopping along the sub-chains and we 
expect average particle currents along the two chains in opposite directions. 
Since the model is particle-number conserving, the currents can be obtained from the continuity equation
$\partial_t \hat{n}_j = \sum_{i}I_{i\to j}-I_{j\to i}$ as a sum of incoming and outgoing terms. In the full model the current along the chain reads
\begin{eqnarray}
     I_{j \to j+2} &=& 2 J\, \textrm{Im}\langle \hat b_{j}^\dagger\hat b_{j+2}\rangle - J g\,  \textrm{Im}\langle \hat b_{j}^\dagger (1-\hat n_{j+1}) \hat b_{j+2}\rangle\nonumber\\
     &&\qquad\pm \sqrt{3} J g\,  \textrm{Re}\langle \hat b_{j}^\dagger (1-\hat n_{j+1}) \hat b_{j+2}\rangle,
     \label{eqn:NNN_current}
\end{eqnarray}
where the upper (lower) sign corresponds to even (odd) $j$ for the upper (lower) chain. %
Note that the appearance of both imaginary and real parts of the correlations is due to the presence of the Peierls phase.
At the same time the average current between the two chains must vanish for periodic boundary conditions. In Fig.~\ref{fig:GS-fidelity}b we have plotted the current $I_{j\to j+2}$ for the upper sub-chain as a function of $g$ and for $\eta=1$. The current along the lower sub-chain is exactly opposite. One notices two interesting features:
\begin{itemize}
    \item[(i)] The direction of the chiral currents reverses when $g$ becomes larger than unity.
    \item[(ii)]  The magnitude of the current is substantially amplified in phase II as compared to phases I and III, with a discontinuous transition.
\end{itemize} 
Furthermore, when $\eta$ is increased the system crosses from the liquid regime into a regime with density order. Consequently, the currents are suppressed with a power law $\sim \eta^{-1}$ and disappear in the limit of large $\eta$ (see \ref{appendix_vortices}).

\section{Origin of different phases in the liquid regime: Mean-field Hamiltonian}\label{section:mean_field_hamiltonian}

In order to get a qualitative understanding of the emergence
of different phases in the liquid regime we first discuss a mean-field approximation.
To this end we express the hard-core bosonic Hamiltonian in terms of fermions using a Jordan-Wigner transformation \cite{Jordan1928} along a zig-zag string (see Fig.~\ref{fig:1D_Model_allTerms}) via
\begin{align}
    \c_i = e^{i\pi\sum_{l<i}\hat{n}_{l}}\b_{i}
\end{align}
Here the $\b_{i}$ are the hard-core boson operators from before and the $\c_{i}$ are fermionic ones.
The transformed Hamiltonian then reads
\begin{eqnarray}
&&\hat{H}=-J\sum_{j}\cd_{j+1}\c_{j}+\cd_{j+2}\c_{j} (1-2\hat{n}_{j+1}) \nonumber\\
&&\quad -2 g J \sum_{j}\cd_{j+2}\c_{j}
\left(e^{\mp \frac{2\pi i}{3}} (1-\n_{j-1})
+e^{\pm \frac{2\pi i}{3}} (1-\n_{j+2}) 
\right)
\nonumber\\
&&\quad -2 g J\sum_{j}\cd_{j+2}\c_{j}
\left[(1-2\hat{n}_{j+1})+2 g\, e^{\pm \frac{4\pi i}{3}}
(1-\n_{j+1})
\right]
\label{eq:H_fermion}\\
&& \quad -2 g J\sum_{j}\cd_{j+3}\c_{j}
\left[e^{\mp \frac{2\pi i}{3}}(1-2\hat{n}_{j+2})(1-\n_{j+1}) +
e^{\pm \frac{2\pi i}{3}}(1-2\hat{n}_{j+1})(1-\n_{j+2})
\right]
\nonumber \\
&& \quad -2 gJ\sum_{j}\cd_{j+4}\c_{j}
(1-2\hat{n}_{j+1})(1-\n_{j+2})(1-2\hat{n}_{j+3})
\nonumber\\
&&\quad -2  \eta g J\sum_{j}\cd_{j}\c_{j}\left[
(1-\hat{n}_{j-1})+(1-\hat{n}_{j+1})+(1-\hat{n}_{j-2})+(1-\hat{n}_{j+2})
\right]+h.c.\nonumber,
\end{eqnarray}
where we have used the identity $e^{\pm i\pi \hat{n}_{j}}=\left(1-2\hat{n}_{j}\right)$.
Now we apply a mean-field approximation, replacing the particle number operators in the higher-order terms by expectation values, assuming a uniform half filling $\expval{\hat{n}_{j}}=0.5$. The corresponding mean-field Hamiltonian  reads
\begin{equation}
\hat{H}_\textrm{MF}=
-J\sum_{j}\cd_{j+1}\c_{j}(1 -  g) 
- gJ\sum_{j}\cd_{j+2}\c_{j}\, e^{\pm \frac{4\pi}{3}i}+h.c.
-2 \eta g J\sum_{j}\cd_{j}\c_{j}.
\label{eq:Hmf}
\end{equation}
One recognizes that the hoppings along the upper and lower sub-chain are complex and contain fixed Peierls phases $\pm 4\pi/3$ of opposite sign, which cannot be gauged away. These phases correspond to a homogeneous effective gauge field, which leads to chiral currents along the upper and lower chain. Furthermore, the direct NN hopping  $j\to j+1$ changes its sign when $g$ becomes larger than unity. 

\begin{figure}[htb]
	\begin{center}
	\includegraphics[scale=0.55]{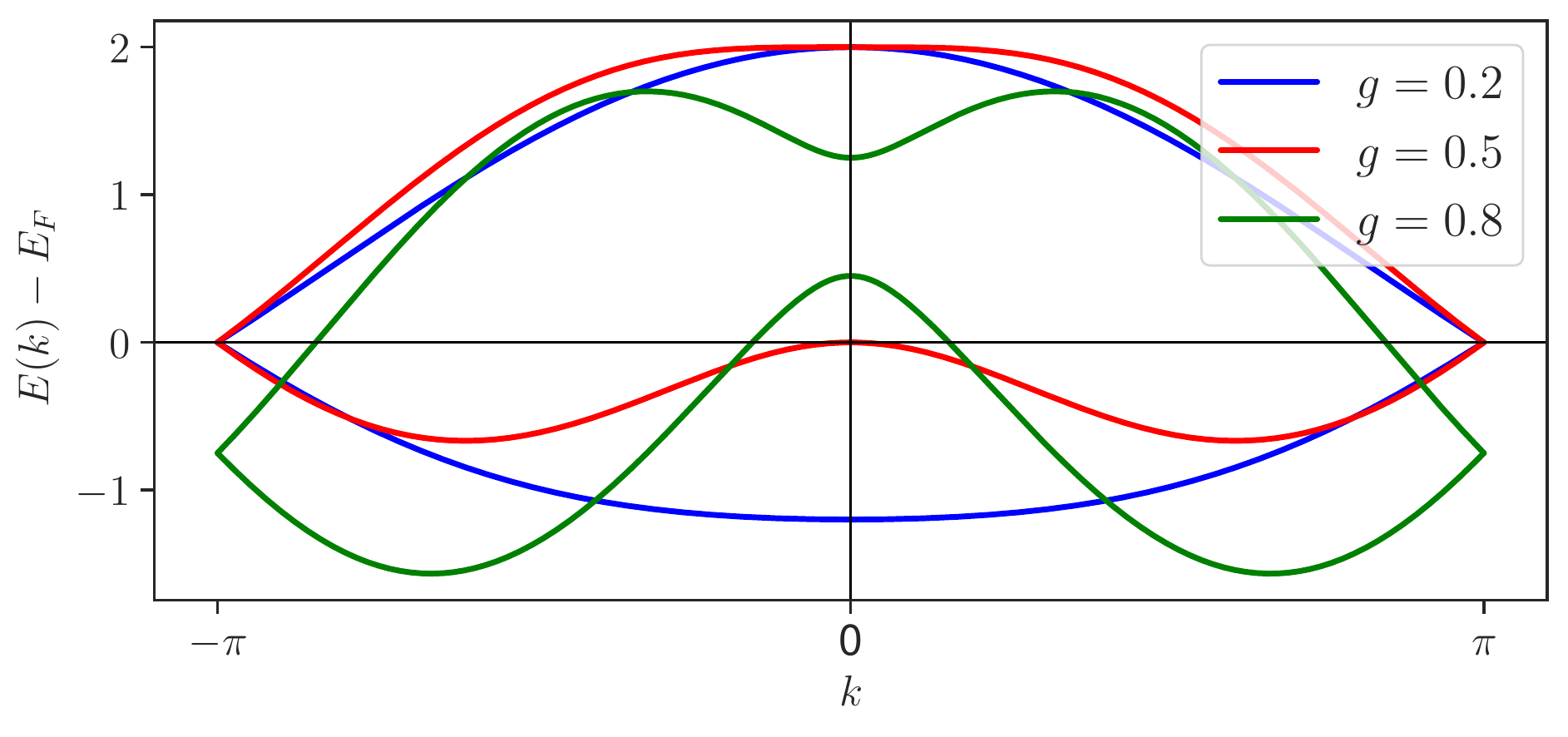}
	\end{center}
	\caption{Single particle spectrum of $\hat H_\textrm{MF}$ for $g=0.2, 0.5, 0.8$ relative to the Fermi energy corresponding to a half-filled system for each value of $g$. One recognizes that for $g\le 0.5$ only the lower band is occupied, while for $g>0.5$ an increasing range of low-momentum states of the lower band is emptied 
	and high-momentum states of the upper band become occupied.
	}
	\label{fig:MF-spectrum}
\end{figure}

We may write $\hat H_\textrm{MF}$ in momentum space
\begin{equation}
    \hat{H}_\textrm{MF}= - J \sum_k \left(\begin{array}{c}\!\! \c_{Ak}^\dagger \!\! \\ \!\! \c_{Bk}^\dagger\!\! \end{array}\right)^{\!\! \top}\!\!
    \left(\begin{array}{cc}  2 g \sin\bigl(k-\frac{\pi}{6}\bigr) & (1-g)(1+e^{-ik}) \\ (1-g) (1+ e^{ik}) & - 2 g \sin\bigl(k-\frac{\pi}{6}\bigr)\end{array}\right)
    \left(\begin{array}{c} \!\!\c_{Ak}\!\! \\ \!\!\c_{Bk}\!\!\end{array}\right),
\end{equation}
where $k\in \left[-\pi,\pi\right]$ is the lattice momentum and the indices $A$ and $B$ denote the upper and lower chain, respectively.
The corresponding spectrum is shown in Fig.\ref{fig:MF-spectrum} where one observes that there is no gap opening between the two bands and thus in mean-field approximation the system is not insulating away from unit filling. For $g< 0.5$ 
only momentum states at the edge of the Brillouin zone touch the Fermi energy. For $g\ge 0.5$, where the system enters phase III, low-momentum states cross the Fermi energy, changing the properties of the many-body ground state.

The nature of the intermediate phase II, in between the superfluid phase I, $g\to0$, and phase III, where second-order hopping processes are dominant, $g\ge 0.5$, cannot be explained within the simple mean-field approximation and requires to consider the interactions included in the full Hamiltonian \eqref{eq:H}.

\begin{figure}[htb]
	\begin{center}
	\includegraphics[width=\columnwidth]{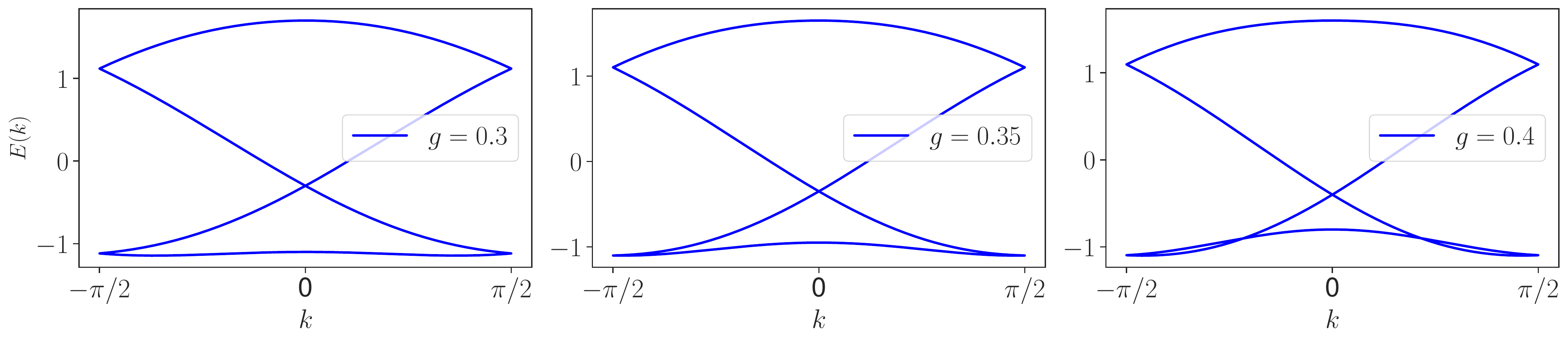}
	\end{center}
	\caption{Folded spectrum of the mean-field Hamiltonian \eqref{eq:Hmf} into half the original Brillouin zone, corresponding to an effective unit cell of 4 lattice sites, where we have set the lattice constant to unity. The latter arises if a density modulation in the two sub-chains is taken into account in the mean-field Hamiltonian. One recognizes a crossing of the lowest sub-bands which occurs for $g\approx 0.366$. This value of $g$ corresponds approximately to the transition point between phase I and phase II.
	}
	\label{fig:MF-spectrum-folded}
\end{figure}

We have seen in section \ref{sec:density-correlations} that the NN density-density interaction drives the system into a density wave of period two in both sub-chains. In the presence of a weak density wave the mean-field Hamiltonian 
gets perturbed. The density-dependent hopping terms are then modulated corresponding to  
a unit cell of four lattice sites. The impact of such a modulation can be understood when folding the 
mean-field spectrum of Fig.~\ref{fig:MF-spectrum} into a reduced Brillouin zone from $-\pi/2$ to $\pi/2$. This is illustrated in Fig.~\ref{fig:MF-spectrum-folded} for values of $g$ between $0.3$ and $0.45$. One recognizes that for $g\ge 0.36$ there is a band crossing of the lower sub-bands, which modifies the nature of the many-body ground state when the density-modulation is taken into account. The emergence of phase II can be attributed to this.

From this mean-field description one naively expects that phase II disappears in the absence of density-density interactions, i.e. for $\eta=0$. Quite surprisingly this is not the case. When calculating the ground-state fidelity of the full model \eqref{eq:H} for $\eta=0$ we still find two peaks at $g\approx 0.5$ and $g\approx 0.375$. This must then be attributed to the presence of the density-dependent complex hopping terms alone. We will return to this point at a later point (\ref{sect:spont_gauge_field_gen}).

\section{Beyond mean-field: Emerging effective gauge fields}\label{section:emerging_gauge}

\begin{figure}[htb]
	\begin{center}
	\includegraphics[width=\columnwidth]{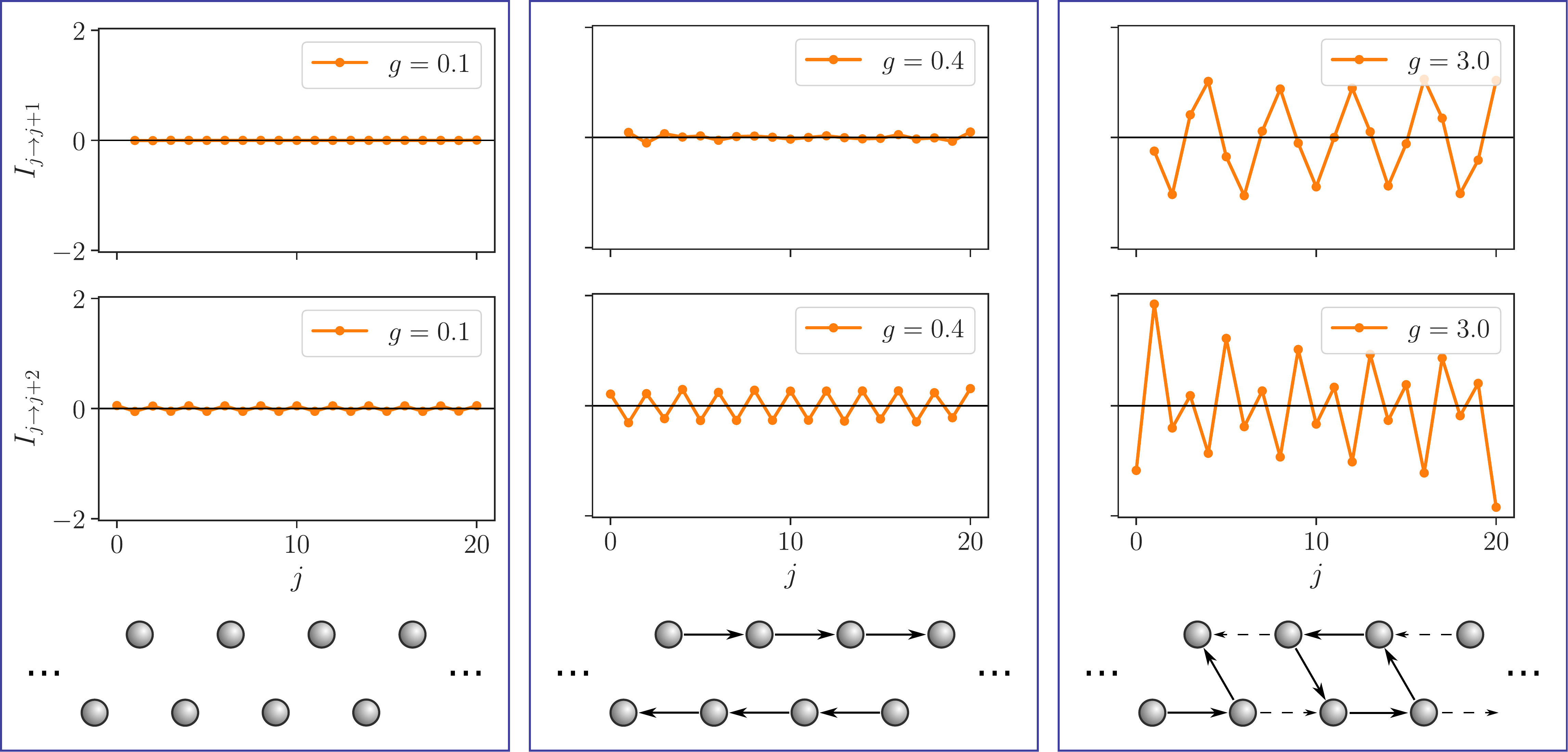}
	\end{center}
	\caption{Currents between the chains (top row) and along the chains (center row) for $\eta=2$ and different values of $g$. One clearly recognizes the formation of a regular pattern of circular currents in the case of large $g$. This pattern is illustrated in the bottom row.
	}
	\label{fig:plaquette_currents}
\end{figure}

To further understand the impact of the density-dependent hopping terms included in the Rydberg-spin Hamiltonian \eqref{eq:H} we once again consider the currents between NN sites between and along the chains. To simplify the methodology we use open boundary conditions (OBC) to select a fixed realization of density-order. In Fig.~\ref{fig:plaquette_currents} we show the corresponding current for OBC spatially resolved, where we have used eq.\eqref{eqn:NNN_current} and
\begin{align}
    I_{j \to j+1} = 
    2 J\, \textrm{Im}\langle \hat{b}_{j}^\dagger\hat{b}_{j+1}\rangle
    &-J g\,  \textrm{Im}\langle \hat{b}_{j}^\dagger (2-\hat{n}_{j+2}-\hat n_{j-1}) \hat{b}_{j+1}\rangle
    \\
    &\pm\sqrt{3} J g\,  \textrm{Re}\langle \hat{b}_{j}^\dagger (\hat{n}_{j+2}-\hat{n}_{j-1}) \hat{b}_{j+1}\rangle.\notag
\end{align}
In the limit $g\to 0$ all currents naturally vanish, as does the effect of the gauge field (left-most column of Fig~\ref{fig:plaquette_currents}). In the middle column we see that for $2g\approx 1$ the currents connecting the sub-chains remain small, whereas the current along the sub-chains shows the mean-field behaviour, inasmuch as we find a homogeneous current moving right in sub-chain A ($j$ even) and moving left in sub-chain B ($j$ odd). However, as soon as the density-dependent hopping become dominant over the direct processes ($g\gg 1$) we observe that both currents feature a periodicity of four sites. We can draw these currents as arrows between the sites (bottom row of Fig~\ref{fig:plaquette_currents}) to see that the current flows around plaquettes of four sites in alternating manner forming a regular array of vortices.

To understand this effect, we rewrite the fermion Hamiltonian \eqref{eq:H_fermion} in the following form 
\begin{eqnarray}
&&\hat{H}=-J \sum_{j}\left[\cd_{j+1}\c_{j} +\cd_{j+2}\c_{j} (1-2\n_{j+1})\right] +h.c.\label{eq:H2}\\
&& - 2 g J\, \sum_j\Bigl[ \cd_{j+1}\c_{j} \, \U_{j+1,j} \bigl(1-\n_{j-1}\n_{j+2}\bigr) 
+ \cd_{j+2}\c_{j}\, U_{j+2,j}
(1-\n_{j+1})+ 
\nonumber\\
&& \quad+\cd_{j+3}\c_{j} \, \U_{j+3,j} \bigl(1-\n_{j+1}\n_{j+2}\bigr) 
- \cd_{j+4}\c_{j}\, (1-2\n_{j+1})(1-\n_{j+2})(1-2\n_{j+3})\Bigr] +h.c.\nonumber\\
&&-2\eta g J \sum_{j}\cd_{j}\c_{j}\,\Bigl[
(1-\hat{n}_{j-1})+(1-\hat{n}_{j+1})+(1-\hat{n}_{j-2})+(1-\hat{n}_{j+2})
\Bigr].\nonumber
\end{eqnarray}
Here we have introduced unitary link operators
\begin{eqnarray}
    \U_{j+1,j} &=& \exp\bigl(i\hat \phi_{j+1,j}\bigr)= -\exp\Bigl(\pm \frac{\pi i}{3} (\n_{j+2}-\n_{j-1})\Bigr),\nonumber\\
    \U_{j+2,j} &=& \exp\bigl(i \hat\phi_{j+2,j}\bigr)= \exp\Bigl(\mp \frac{\pi i}{3}\Bigr),\label{eq:gauge}\\
    \U_{j+3,j} &=& \exp\bigl(i\hat \phi_{j+3,j}\bigr) = -\exp\Bigl(\pm \frac{2\pi i}{3} (\n_{j+1}-\n_{j+2})\Bigr).\nonumber
\end{eqnarray} 
where the upper (lower) sign corresponds to even (odd) indices $j$. Thus besides the direct hopping amplitudes 
at rate $J$ there are complex hopping terms proportional to $Jg$ corresponding to an effective gauge field. 
In contrast to lattice models describing particles coupled to a \emph{fixed} 
background field, such as the Harper-Hofstadter model, the effective gauge fields are here operator valued.
Only $\U_{j+2,j}$ corresponds to a fixed, classical gauge field.

Despite the fact that the gauge fields causing the
operator-valued Peierls phases are dynamical quantities, there is no additional conserved charge associated with them and the gauge freedom is the same as for a classical
background field. 
In the case of a classical background field, the unitary 
link operators are just exponential phase factors 
\begin{equation}
U_{lj} = \exp\left\{i\int_j^l \!\! d\vec{r}\cdot \vec{A}\right\},
\end{equation}
where $\vec{A}$ is the external, classical vector potential. Under a ${\sf U}(1)$ gauge transformation of the classical field, the link operator transform as
\begin{equation}
 U_{lj}\,  \to \, U_{lj}^\prime=
  \exp\left\{i\int_j^l \!\! d\vec{r}\cdot (\vec{A} + \vec{\nabla}\alpha)\right\}
  =
  \exp\left\{i\int_j^l \!\! d\vec{r}\cdot \vec{A} + i\bigl(\alpha_{l}-\alpha_{j}\bigr)\right\},\label{eq:gaugetrafo}
\end{equation}
which can be compensated by a unitary transformation of the fermion operators, 
$e^{-i\alpha_j \n_j} \, \hat c_j \, e^{i\alpha_j \n_j} = \hat c_j \, e^{i\alpha_j}$,
such that the Hamiltonian is invariant under the total transformation. 
We should remember that only the longitudinal component of the vector potential
$\vec{A}=\vec{A}_\parallel+\vec{A}_\perp$ changes under a gauge transformation, while the transversal part $A_\perp$ remains the same.
Upon inspecting eq.\eqref{eq:gauge} one recognizes that the operator-valued contributions $\hat {\vec A}$ to the gauge field  are transverse:
\begin{eqnarray}
\hat \phi_{l,j} = \int_j^l\!\! d{\vec{r}} \cdot(\vec{A}+ \hat {\vec{A}})
= \int_j^l\!\! d{\vec{r}} \cdot (\vec{A}_\parallel +\hat {\vec{A}}_\perp)
\end{eqnarray}
 To see this we note that the operator terms in eqs.\eqref{eq:gauge} are given by differences of occupation numbers at lattice sites left and right of the hopping path, which is the lattice equivalent of
\begin{eqnarray}
\hat {\vec{A}} \sim  {\vec{\nabla}} \times \n(x,y) {\vec{e}}_z = \left(\frac{\partial \n}{\partial y},-\frac{\partial \n}{\partial x},0\right), \quad \textrm{and thus}\quad \vec\nabla\cdot \hat {\vec A}=0. 
\end{eqnarray}
The Peierls phases $\hat \phi_{l,j}$ themselves have no direct physical relevance as a classical gauge transformation of the longitudinal component, eq.\eqref{eq:gaugetrafo}, leads to  $\hat \phi_{l,j} \to \hat \phi_{l,j} + \alpha_l-\alpha_j$.
Quantities that are invariant under these gauge transformations are plaquette fluxes, e.g. in a rhombus of lattice sites $j,j+1,j+2,j+4$ consisting of two adjacent triangles, $\hat \Theta_j^\square$ (mod $2\pi$).
$\hat \Theta_j^\square$ contains a fixed classical component
$\Phi_c^\square = 2\pi/3$ and a quantum component $\hat \Phi_j^\square$
\begin{eqnarray}
\hat \Theta_j^\square &=& \pm\Bigl(\hat\phi_{j+2,j} + \hat \phi_{j+3,j+2} - \hat \phi_{j+3,j+1} - \hat \phi_{j+1,j} \Bigr),\label{eq:flux}\\
&=& -\frac{\pi}{3} \Bigl(\n_{j+2}+\n_{j+1} -\n_{j-1}-\n_{j+4}\Bigr) +\frac{2\pi}{3}= \hat \Phi_j^\square + \Phi_c^\square,\nonumber
\end{eqnarray}
where the upper (lower) sign correspond to even (odd) sites $j$. Note that we have defined the sign of the fluxes 
by a clockwise
sequence of sites in the $+z$ direction in Fig.\ref{fig:1D_Model_allTerms}, i.e. for even $j$: $j\to j+2\to j+3\to j+1 \to j$ and for odd $j$: $j\to j+1\to j+3\to j+2\to j$.
In the absence of density modulations the quantum gauge fields
$\hat \Phi_j^\square$ have all a vanishing magnitude and one recovers the mean-field behaviour. This changes however if density correlations are taken into account. Then the plaquette fluxes are modulated and they induce currents between the two chains
$I_{j\to j+1}$ and modulate the currents $I_{j\to j+2}$ along the chains, as seen in Fig~\ref{fig:plaquette_currents}.

\subsection{Dynamical gauge fields driven by density-density interactions}
\label{Sect:density-driven-gauge}
As shown in Fig.\ref{fig:density-correlations}, density-density interactions drive the many-body ground state
towards an ordered state with a density wave for large values of $\eta$. For the ground states \eqref{eq:states} one would find e.g. in state $\vert \psi_0\rangle =\dots \vert 0\rangle_0\vert 1\rangle_1\vert 1\rangle_{2}\vert 0\rangle_{3}
\vert 0\rangle_{4}\vert 1\rangle_{5}\dots$
\begin{eqnarray*}
\langle\hat \Phi_0^\square\rangle &=& \frac{2\pi}{3},\qquad
\langle\hat \Phi_1^\square\rangle = 0,\qquad
\langle\hat \Phi_2^\square\rangle = -\frac{2\pi}{3},\qquad
\langle\hat \Phi_3^\square\rangle = 0,\quad\textrm{etc.}
\end{eqnarray*}
Thus the total flux of the gauge field oscillates around the classical value $\Phi_c^\square =2\pi/3$. 

\begin{figure}[htb]
	\begin{center}
	\includegraphics[width=0.48\columnwidth]{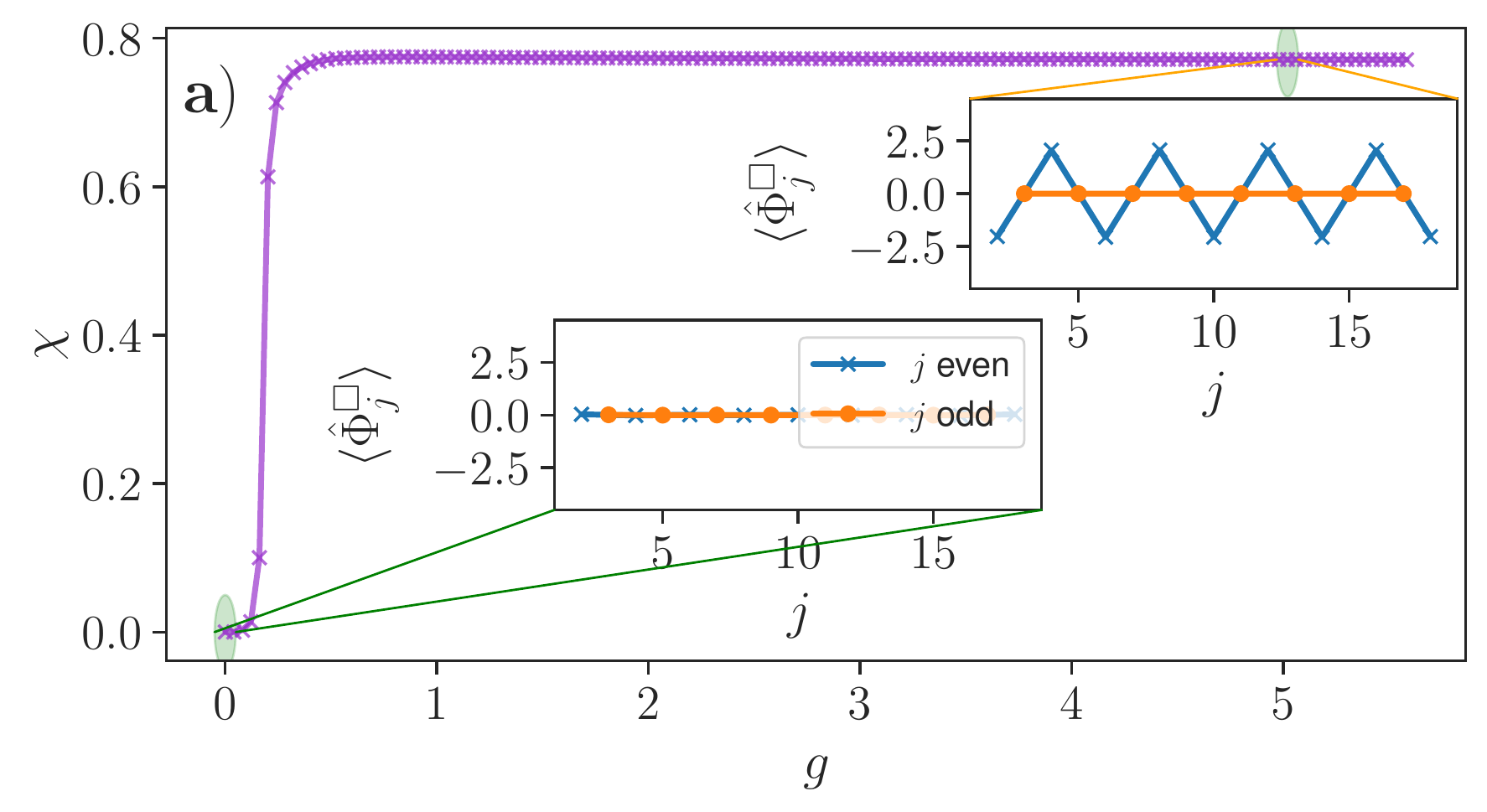}
	\includegraphics[width=0.48\columnwidth]{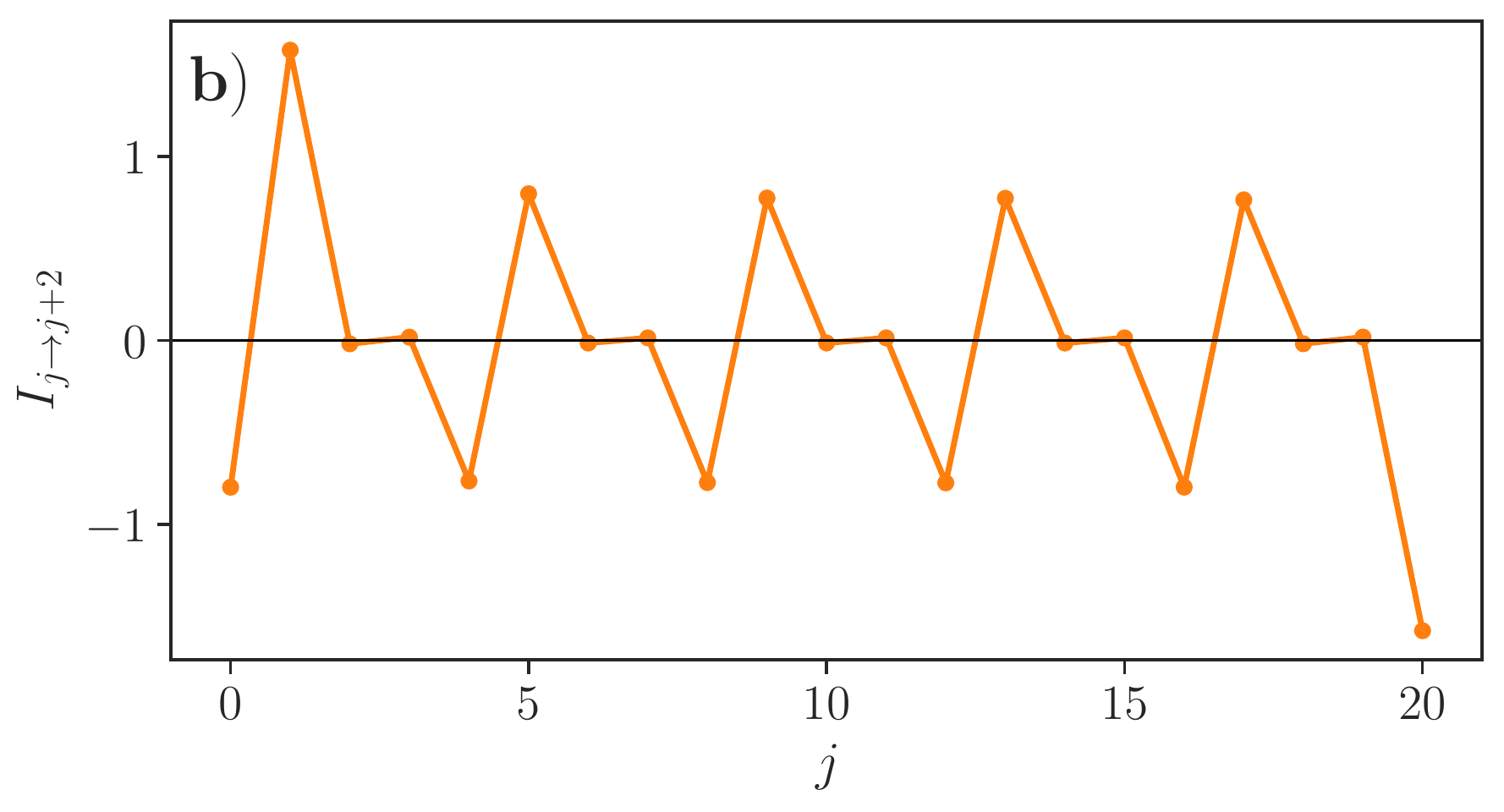}
	\end{center}
	\caption{(a) Order parameter of flux lattice $\chi$ in the ordered phase for $\eta=10$, excluding the classical flux of $2\pi/3$. Zoom-ins show the plaquette fluxes $\hat{\Phi}_{j}^\square$ in the highlighted region. 
	(b) Currents along the sub-chains in the orderd phase for $\eta=10$ and OBC. Although the currents are suppressed with $1/\eta$, the long-range ordered pattern is clearly visible.
	}
	\label{fig:emerging_gauge_field}
\end{figure}

To describe the transition to the flux-ordered phase 
in a system with open boundary conditions,
we introduce the order parameter
\begin{align}
    \chi = \abs{\frac{1}{L}\sum_{j=2n}\left(-1\right)^{n}\left(\expval{\hat{\Phi}_{j}^\square}+\expval{\hat{\Phi}_{j+1}^\square}\right)}.
    \label{eqn:chi_order_param}
\end{align}
In this expression, we exclude the first and last plaquette at the edges of the system to avoid boundary effects. In Fig.\ref{fig:emerging_gauge_field}a we show $\chi$ as function of $g$ in the regime of large $\eta$. 
We recognize a transition into a phase with long-range order of the flux lattice. The origin of the regular pattern of staggered effective fluxes is simply the density order generated by the
repulsive interaction proportional to $\eta$. 
Since in the ordered phase (IV) first-order correlations are not strictly zero but are suppressed proportional to $\eta^{-1}$ (see Appendix),
the flux pattern induces a lattice of current vortices similar to those in Fig.\ref{fig:plaquette_currents}. In 
Fig.\ref{fig:emerging_gauge_field}b we have plotted the local current between the two sub-chains inside the ordered phase IV for open boundary conditions.

\subsection{Spontaneous gauge field generation}\label{sect:spont_gauge_field_gen}

When $\eta$ is reduced and below the critical value of the
transition to the ordered phase, density correlations show a power-law decay as function of distance. In this case the order parameter $\chi$ vanishes in the thermodynamic limit, but the local flux $\hat \Phi_j^\square$ shows non-vanishing fluctuations, i.e. $\expval{\bigl(\hat \Phi_j^{\square}\bigr)^{2}}\ne 0$ and liquid-like
spatial correlations.

\begin{figure}[htb]
	\begin{center}
	\includegraphics[width=\columnwidth]{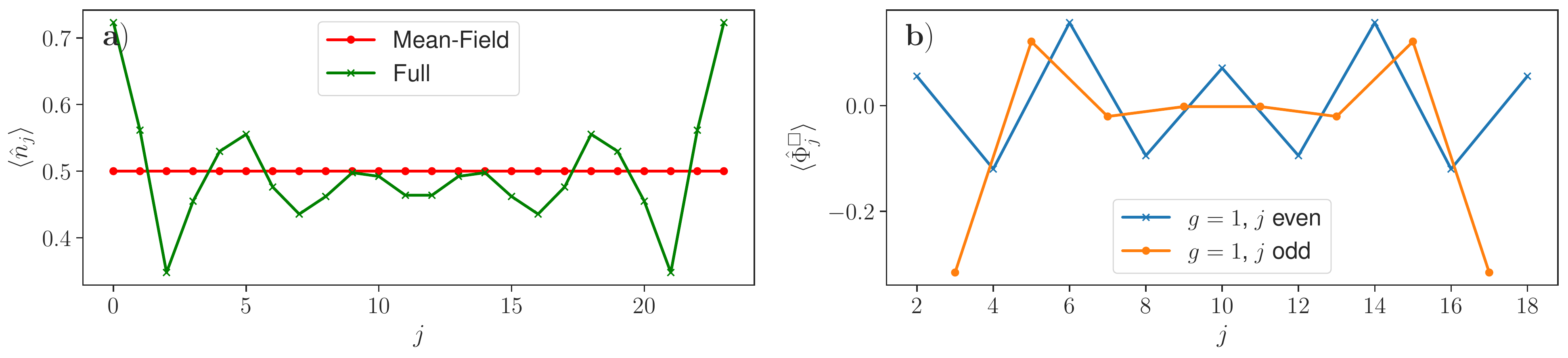}
	\end{center}
	\caption{
	    a) Average density in the mean-field and full models in the absence of density-density repulsion, i.e. $\eta=0$, and $g=1$, where the sub-chains decouple in the mean-field case. b) Full model: Spatially resolved flux of quantum gauge field $\langle\hat{\Phi}^{\square}_j\rangle$ for finite system ($L=24$) with OBC for the same parameter. Lattice sites directly at the boundary are cut off. One clearly recognizes a spontaneous generation of the gauge field with liquid-like spatial correlations.
	}
	\label{fig:density_order_MF_vs_full_eta=0}
\end{figure}

However, when the density-density interaction is switched off entirely, i.e. for $\eta =0$, one naively expects no density modulations in the absence of density-dependent complex hoppings. In fact due to the triangular lattice structure Friedel oscillations do lead to density modulations in a finite system with open boundary conditions even in the mean-field limit. Only at $g=1$, where the mean-field Hamiltonian 
\eqref{eq:Hmf} separates into two independent chains, the corresponding density distribution is flat, see 
Fig.\ref{fig:density_order_MF_vs_full_eta=0}a, and we expect the quantum gauge field to be in a vacuum state.
We note that the classical Peierls phases can be gauged away in this  case.
The full model does however show density modulations, see  Fig.\ref{fig:density_order_MF_vs_full_eta=0}a, which are entirely caused by the density-dependent complex hopping terms. Thus there is a spontaneous generation of the gauge field with 
 non-vanishing amplitude of $\langle\bigl(\hat \Phi_j^\square\bigr)\rangle$ 
and fluxes that form an oscillating pattern for open boundary conditions (OBC). The oscillations decay away from the
boundary, however, showing that the flux correlations are liquid-like and there is no long-range order, in contrast to Fig.\ref{fig:emerging_gauge_field}a for $\eta =10$.

\subsection{Gutzwiller Ansatz}
\begin{figure}[htb]
	\begin{center}
	\includegraphics[width=0.55\columnwidth]{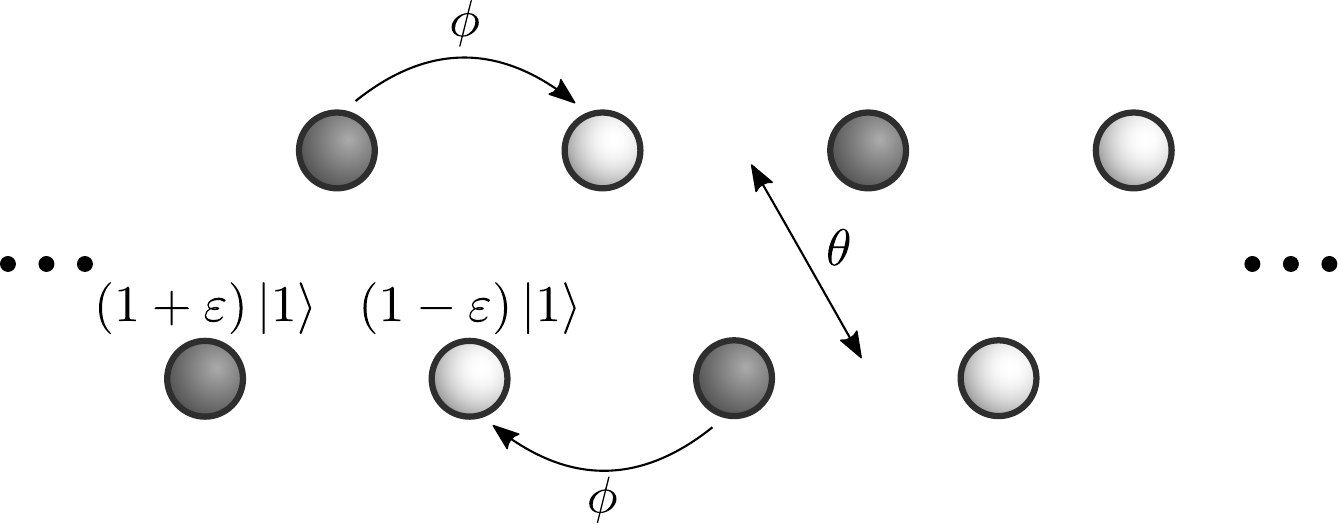}
	\end{center}
	\caption{
	    Schematic representation of the Gutzwiller ansatz. The darker (brighter) sites correspond to an increased (decreased) excited state occupation. The arrows symbolize the different complex phases.}
	\label{fig:Gutzwiller_ansatz}
\end{figure}

%
The spontaneous generation of a non-zero expectation value of the staggered gauge field results from the competition between an increase of kinetic energy 
of direct hopping processes on the one hand and a lowering of energy by increase of the imaginary second-order hopping amplitudes on the other. 
To see this, we performed a variational approach to determine the ground state using a Gutzwiller ansatz.
Guided by the numerical results we assume a 
unit cell of four sites and write the following ansatz wavefunction, illustrated in Fig.\ref{fig:Gutzwiller_ansatz} 
\begin{align}
    \ket{\psi}&=\frac{1}{\mathcal{N}}\prod_{j=4n}\ket{\psi_{j}}\ket{\psi_{j+1}}\ket{\psi_{j+2}}\ket{\psi_{j+3}}
    \\
    \ket{\psi_{j}}&=(1-\varepsilon)\ket{0}+(1+\varepsilon)\mathrm{e}^{\mathrm{i}(\theta-\phi)}\ket{1}
    \\
    \ket{\psi_{j+1}}&=(1+\varepsilon)\ket{0}+(1-\varepsilon)\mathrm{e}^{\mathrm{-i}(\theta-\phi)}\ket{1}
    \\
    \ket{\psi_{j+2}}&=(1+\varepsilon)\ket{0}+(1-\varepsilon)\mathrm{e}^{\mathrm{i}(\theta+\phi)}\ket{1}
    \\
    \ket{\psi_{j+3}}&=(1-\varepsilon)\ket{0}+(1-\varepsilon)\mathrm{e}^{\mathrm{-i}(\theta+\phi)}\ket{1}.
\end{align}
Here we use three parameters to accommodate the physics of the model. The parameter $\varepsilon$ accounts for the observed density modulation, $\theta$ distinguishes the two sub-chains and $\phi$ accounts for the complex phase that is picked up when a particle hops along the same sub-chain (to the right for the upper chain, to the left for the lower one). 

\begin{figure}[htb]
	\begin{center}
	\includegraphics[width=\columnwidth]{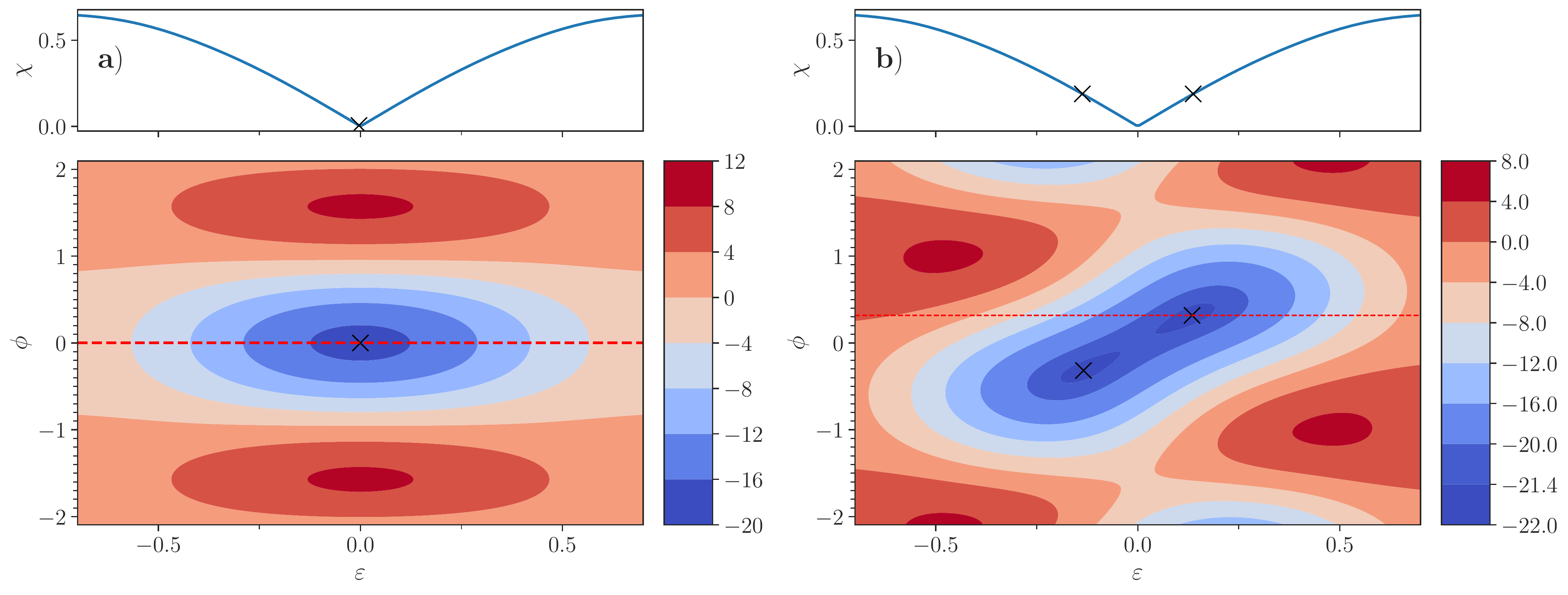}
	\end{center}
	\caption{
	    \textit{Bottom:} Contour-Plot of the Gutzwiller-energies for $L=16, \eta=0$. a): $g=0.001$, b): $g=1$. Shown is a cut through the 3-dimensional parameter space at the optimal values of $\theta$, respectively ($\theta=0$ and $\theta=\pi/2$). Crosses mark the (degenerate) minima. \textit{Top:}  Order parameter $\chi$ from \eqref{eqn:chi_order_param} for values along the red dotted lines in the bottom figures.}
	\label{fig:Gutzwiller_double_plot}
\end{figure}

Using a standard optimization algorithm to minimize the variational energy, we obtain the following results for the three parameters (up to numerical precision) for $L=16$, $\eta=0$ and $g=1$:
\begin{align}
    \varepsilon_{opt} \approx 0.134
    &&
    \theta_{opt} \approx \frac{\pi}{2}
    &&
    \phi_{opt} \approx 0.317
\end{align}
For the case of very small density-dependent hopping, $g\to 0$, all three optimal parameter values are found to be zero as in this case only direct hopping processes survive.
In Fig.~\ref{fig:Gutzwiller_double_plot} we show the expectation value of the ground-state energy in the Gutzwiller state, where we have fixed the third parameter $\theta$ to its optimal values, $\theta=0$ ($g=0.001$) and $\theta=\pi/2$ ($g=1$). For small density-dependent hopping the alternating currents quantified by $\chi$ vanish. For $g=1$ however there exist two degenerate minimum energy configurations point-symmetric to zero which represent the two different chain configurations and feature non-vanishing gauge fields, as indicated by the value of $\chi$ for the horizontal cut in Fig.~\ref{fig:Gutzwiller_double_plot}. 

\begin{figure}[htb]
	\begin{center}
	\includegraphics[width=0.6\columnwidth]{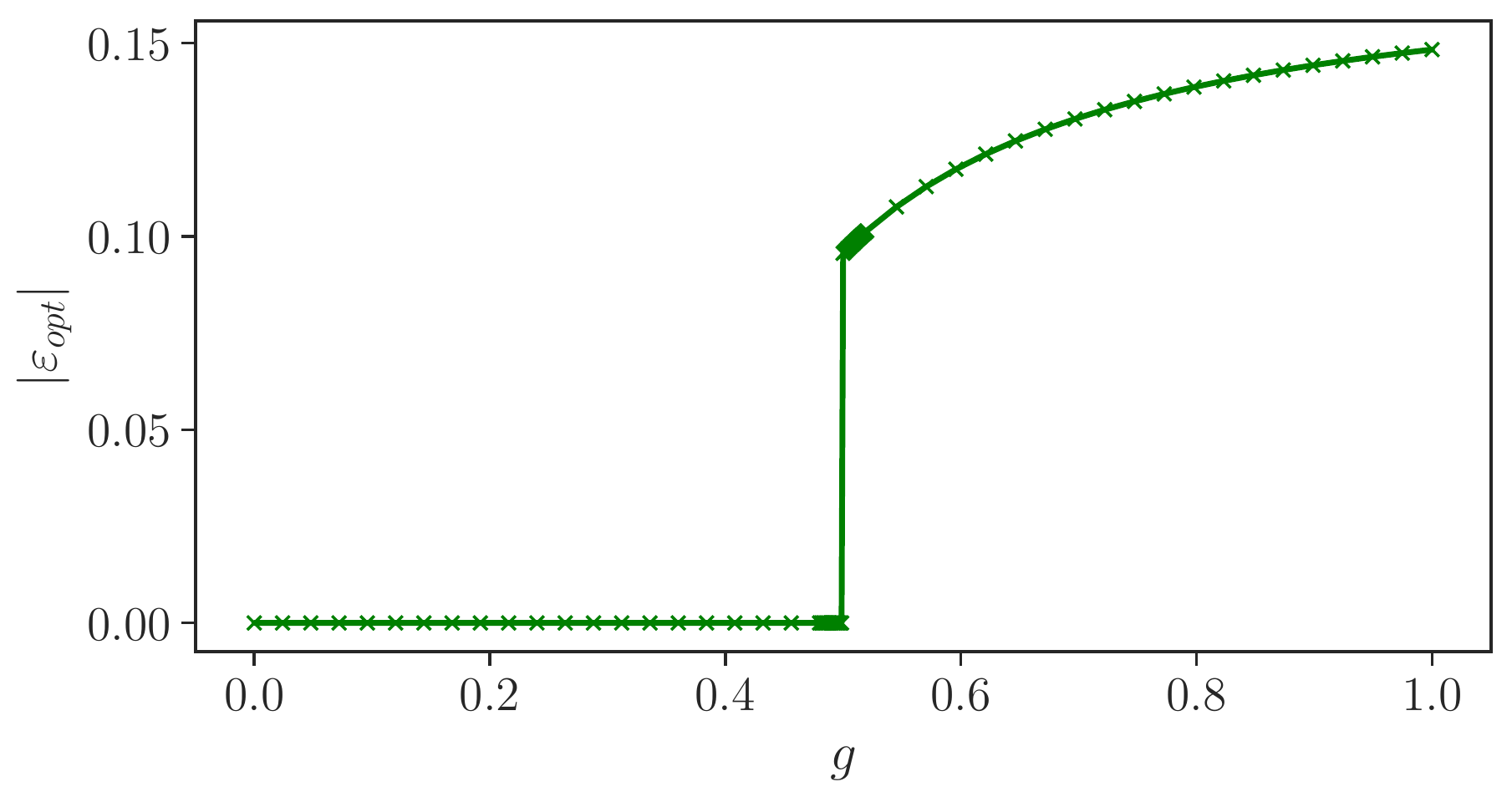}
	\end{center}
	\caption{Density modulation $\varepsilon$, which is linked to the gauge field amplitude, as function of the strength of second-order hopping processes $g$ in Gutzwiller approximation for $\eta=0$. The critical value agrees qualitatively with the transition to the gauge-field dominated regime III in Fig.\ref{fig:phase-diagramm}.}
	\label{fig:Gutzwiller-epsilon}
\end{figure}

We note that the Gutzwiller ansatz cannot capture a liquid-type phase with short-range correlations of the gauge field and thus a finite value of the order parameter $\chi$ is obtained. It is instructive to consider the optimum value of the density imbalance, characterized by the variational parameter $\varepsilon$ as function of the strength $g$ of second-order hopping processes. This is shown in Fig.\ref{fig:Gutzwiller-epsilon}. 
One recognizes that within the Gutzwiller approach there is a phase transition at $g=0.5$ from a phase with vanishing gauge field to a phase with a finite amplitude. Below the critical value of $g$ the energy increase due to a density wave outweights the energy decrease due to the gauge-field mediated hopping. Above the critical value the situation is reversed.

\section{Possible experimental signatures of non-trivial liquid phases}
\label{sec:exp_realization}

\begin{figure}[htb]
	\begin{center}
	\includegraphics[width=0.95\columnwidth]{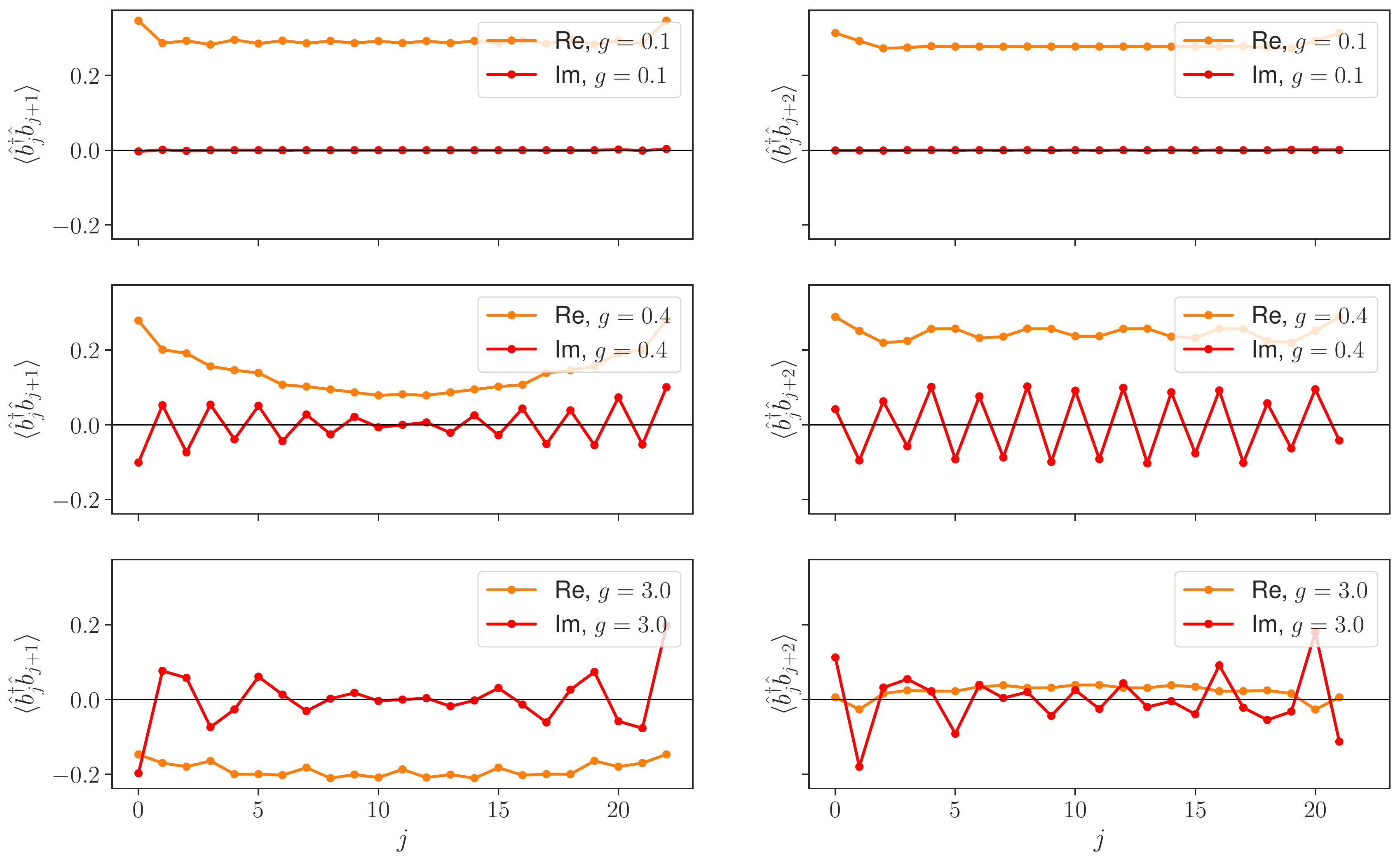}
	\end{center}
	\caption{First-order correlations $\langle\hat{b}_{k}^{\dagger}\hat{b}_{l}\rangle$ for open boundary conditions,$\eta=1$ and system size $L=24$. Left: inter-chain correlations between sites $j$ and $j+1$, right: intra-chain correlations between sites $j$ and $j+2$. From top to bottom row: $g=0.1, 0.4, 3.0$.}
	\label{fig:FirstOrderCorr}
\end{figure}

While vortex currents of Rydberg excitations are difficult to measure in experiments, first order correlations can be accessed. Applying a global microwave pulse to all Rydberg atoms allows to rotate all spins formed by states $\vert 0\rangle$ and $\vert 1\rangle$, see Fig. 
\ref{fig:same_species_all}c, from the $x$- or $y$-direction on the Bloch-sphere onto the $z$ axis. A subsequent spatially resolved measurement of correlations of Rydberg excitations thus yields $xx$ or $yy$ spin correlations, from which one can obtain the real part of first-order boson correlations
\begin{equation}
    \mathrm{Re}\bigl[\langle \hat b_k^\dagger \hat b_l\rangle\bigr] = \langle \hat\sigma_k^x\hat\sigma_l^x\rangle +\langle\hat\sigma_k^y\hat\sigma_l^y\rangle.
\end{equation}
Using laser beams addressing specific atoms and a global microwave pulse, it is possible to apply spin rotations around different axes to different Rydberg atoms. In this way anisotropic spin correlations can be measured
which give access also to the imaginary part of first-order correlations between bosons at different lattice sites
\begin{equation}
    \mathrm{Im}\bigl[\langle \hat b_k^\dagger \hat b_l\rangle\bigr] 
    =
    \langle\hat\sigma_k^y\hat\sigma_l^x\rangle - \langle \hat\sigma_k^x\hat\sigma_l^y\rangle.
\end{equation}
As shown in the phase diagram Fig.\ref{fig:phase-diagramm} the non-trivial liquid phases II and III can be distinguished from the trivial phase I by first-order correlations of neighboring lattice sites between the sub-chains, $\langle \hat b_j^\dagger \hat b_{j+1}\rangle$, and along the sub-chains, $\langle \hat b_j^\dagger \hat b_{j+2}\rangle$.
In phase I the real part of both is positive and there is no imaginary component, indicating the absence of currents, see top row of Fig.\ref{fig:FirstOrderCorr}. In phase II the imaginary parts of both correlations feature an oscillating pattern, while the real parts remain positive with reduced amplitudes, see middle row of Fig.\ref{fig:FirstOrderCorr}. The nearly homogeneous oscillation in the right column is a signature of the constant chiral currents along the upper and lower sub-chain. Finally in phase III the real part of the inter-chain correlation $\langle \hat b_j^\dagger \hat b_{j+1}\rangle$ flips its sign and the imaginary parts attain a more complex pattern reminiscent of vortex currents.
To summarize, using the technique of global (and local) microwave pulses and subsequent space-resolved measurement of Rydberg excitations allows to experimentally confirm the non-trivial liquid regimes of the model.

\section{Summary and Conclusion}
\label{sec:summary}
We studied the effects of a density-dependent, complex hopping of Rydberg excitations arising from second-order processes to the many-body ground state of a one-dimensional zig-zag-chain at half filling.
 The strength of the second-order hopping  can be controlled by microscopic parameters of the Rydberg atoms and the system can be tuned from a regime where these processes are negligible to one where they dominate. 
Second order processes also lead to density-density interactions, which are modified by additional off-resonant van-der-Waals couplings between Rydberg atoms. The competition between direct and second-order hopping processes as well as density-density repulsion leads to a rich ground-state phase diagram.
Most importantly the complex valuedness of the second-order hopping amplitudes breaks time-reversal symmetry and, as shown in the present paper, leads to two types of effective gauge fields. The first one is equivalent to a classical vector potential. It is worth noting that this field emerges directly from the microscopic physics of Rydberg atoms and is very different from other mechanisms that have been suggested for effective gauge fields for neutral atoms. This classical gauge field can be employed for example to realize topological band structures. Even more interesting is the second kind of gauge field which arises from the density-dependence of the
second-order hopping. It is a quantum field linked to density modulations. Thus we here encounter a \textit{dynamical} quantum gauge field. While based on
the mechanism discussed here it is possible to implement true lattice gauge theories (publication in preparation), the operator-valued field in the present context is a transversal vector potential. It is thus gauge invariant and does not lead to additional conserved local charges. 
Depending on the values of second order hopping and density-density interaction we identified four distinct phases. An insulating phase with long-range density order in a repulsion-dominated regime, a superfluid for weak interactions, and two nontrivial liquid phases where density repulsion is weak but second order hopping dominates over direct hopping. In the insulating phase long-range density correlations induce
an oscillatory lattice of fluxes of the quantum gauge field, which in turn leads to a regular pattern of local vortex currents. Surprisingly, even in the complete absence of density-density interactions the quantum gauge field attains a non-vanishing value with liquid-like correlations. We must interpret this as a spontaneous
creation of the gauge field by breaking of the lattice translational symmetry.
\\
Our findings demonstrate that systems of Rydberg atoms can give rise to a broad spectrum of nontrivial interaction processes and offer new and 
experimentally accessible approaches to study dynamical gauge fields. Additionally, even topologically non-trivial states are realizable using this Rydberg system, as we demonstrate in a Haldane-type model on a honeycomb lattice \cite{Ohler2021a}.

\subsection*{Acknowledgement}
S.O., M.K. and M.F. gratefully acknowledge financial support from the DFG through SFB TR 185, project number 277625399. A.B. acknowledges financial support from the European Union's Horizon 2020 research and innovation program under grant agreement no. 817482 (PASQuanS). A.B. and H.P.B. acknowledge financial support from the French-German collaboration for joint projects in NLE Sciences funded by the Deutsche Forschungsgemeinschaft (DFG) and the Agence National de la Recherche (ANR PRCI, project RYBOTIN).
\\
The authors would like to thank David Petrosyan for very insightful discussions.

\appendix
\section{Microscopic Derivation}\label{sect:App_MicroDerivation}
\newcommand{\iup}{\mathrm{i}}
In this section we provide a microscopic derivation of  Hamiltonian \eqref{eqn:basic_triangle_hamiltonian}.
\\
We start from the static dipole-dipole interaction between two atoms
\begin{align}\label{eqn:App_V_ij_finalform}
\hat{V}_{ij}&=\frac{1}{4\pi\epsilon_{0}R_{ij}^{3}}\left[
\hat{d}^{z}_{i}\hat{d}^{z}_{j}+\frac{1}{2}\left(\hat{d}^{+}_{i}\hat{d}^{-}_{j}+\hat{d}^{-}_{i}\hat{d}^{+}_{j}\right)-\frac{3}{2}
\left(
\hat{d}^{+}_{i}\hat{d}^{+}_{j}\mathrm{e}^{-2i\phi_{ij}}+\hat{d}^{-}_{i}\hat{d}^{-}_{j}\mathrm{e}^{2i\phi_{ij}}
\right)
\right],
\end{align}
where $R_{i,j}=\abs{\vec{r}_{i,j}}$ and $\phi_{ij}$ refer to the interatomic distance and angle in polar coordinates, respectively. Additionally, the operators $\hat{d}^{\pm}_{i}=\mp\frac{1}{\sqrt{2}}(\hat{d}^{x}_{i}\pm i\hat{d}^{y}_{i})$ represent ladder operators raising and lowering the angular momentum of the $i$-th Rydberg atom. The total Hamiltonian of the three-site setup shown in Fig.~\ref{fig:same_species_all} then reads
\begin{align}\label{eqn:three_site_total_H}
    \hat{H}=\sum_{i=1}^{3}\left(\omega\ketbra{1}_{i}+\left(\omega+\Delta\right)\ketbra{+}_{i}\right)+\sum_{i\neq j}\hat{V}_{ij}.
\end{align}
Here, $\omega$ refers to the energy difference between the states $\ket{1}$ and $\ket{0}$, while $\Delta$ denotes the detuning of the state $\ket{+}$ shown in Fig.~\ref{fig:same_species_all}. The subscript of the projectors indicates the respective atom.
In deriving the spin excitation transport in the triangle of Rydberg atoms we only need to consider the case of one and two excitations in the system, where we define an atom to be excited if it occupies either the state $\ket{1}$ or $\ket{+}$. This definition of occupation differs from the main text, but is convenient for the derivation.
\\
For the case of a single excitation, we adiabatically eliminate the state $\ket{+}$, which is detuned by $\Delta$ and thus will only be virtually excited. This leads to 
a $3\times 3$ matrix describing the effective dipole-dipole interaction of three two-level systems with a single excitation (in state $\ket{1}$ or $\ket{+}$)
\begin{align}\label{eqn:H_eff_triangle_3x3matrix}
\hat{H}_{eff}=
\begin{pmatrix}
0 & \bra{100}\hat{H}\ket{010} & \bra{100}\hat{H}\ket{001} \\
\bra{010}\hat{H}\ket{100} & 0 & \bra{010}\hat{H}\ket{001} \\
\bra{001}\hat{H}\ket{100}  & \bra{001}\hat{H}\ket{010} & 0 \\
\end{pmatrix}
-\frac{1}{\detuning}
\begin{pmatrix}
U_{1} & h_{1\rightarrow 3\rightarrow 2} & h_{1\rightarrow 2\rightarrow 3}
\\
h_{2\rightarrow 3\rightarrow 1} & U_{2} & h_{2\rightarrow 1\rightarrow 3}
\\
h_{3\rightarrow 2\rightarrow 1} & h_{3\rightarrow 1\rightarrow 2} & U_{3}
\end{pmatrix},
\end{align}
where $\vert 100\rangle$ denotes the state where the three atoms are in states $\vert 1\rangle$, $\ket{0}$, $\ket{0}$, respectively. In this calculation we neglected all terms of higher order in $1/\Delta$.
In equation \eqref{eqn:H_eff_triangle_3x3matrix}, the symbol $h_{i\rightarrow j\rightarrow k}$ refers to the indirect hopping process going from the $\ket{1}$ state at site $i$ via the $\ket{+}$ state at site $j$ to  $\ket{1}$ at site $k$:
\begin{align}
h_{1\rightarrow 2\rightarrow 3}=
\bra{0,0,1}\hat{H}\ket{0,+,0}\bra{0,+,0}\hat{H}\ket{1,0,0}.
\end{align}
Therefore, $h_{i\rightarrow j\rightarrow k}$ describes exactly the process shown in Fig.~\ref{fig:IndirectProcess_visual} in the main text. Similarly, the symbol $U_{i}$ denotes the virtual processes where the initial site $i$ and the final site $k$ are identical, e.g. for $i=1$:
\begin{align}
U_{1} = h_{1\rightarrow 2\rightarrow 1}+h_{1\rightarrow 3\rightarrow 1}.
\end{align}
A key aspect is the fact that the latter $1/\Delta$ term in \eqref{eqn:H_eff_triangle_3x3matrix} only occurs within the single-excitation manifold.
E.g. if site $2$ is occupied, the matrix element vanishes
\begin{align}
\bra{0,1,1}\hat{H}\ket{0,+,0}\bra{0,+,0}\hat{H}\ket{1,1,0}=0,
\end{align}
and only the direct dipole-dipole flopping process between site $1$ and site $3$ remains (first term in \eqref{eqn:H_eff_triangle_3x3matrix}). The identical consideration holds for the terms $U_{i}$. As a result, both the hopping amplitudes and the nearest neighbor interaction terms are density-dependent. We can make this effect explicit by writing down the Hamiltonian using spin operators. We define the energy scale 
\begin{align}\label{eqn:def_energy_scale}
J = -\bra{0,1,0}\hat{H}\ket{0,0,1} = -\bra{0,1,0}\hat{V}_{1,2}\ket{1,0,0}
\end{align}
of the direct hopping process over the distance $R$ as shown in Fig.~\ref{fig:same_species_all}. The explicit evaluation of these matrix elements can be done using Wigner-symbols. For this, we reduce the expectation values of the interaction Hamiltonian \ref{eqn:def_energy_scale} to the dipole operators
\begin{align}
    \bra{0,1,0}\hat{V}_{1,2}&\ket{1,0,0}
\\
=&
\frac{1}{8\pi\epsilon_{0}R_{1,2}^{3}}
\bra{0,1,0}
\left(\hat{d}^{+}_{1}\hat{d}^{-}_{2}+\hat{d}^{-}_{1}\hat{d}^{+}_{2}\right)
\ket{1,0,0}
\\
=&
\frac{1}{8\pi\epsilon_{0}R^{3}}
\bra{0,1,0}
\hat{d}^{-}_{1}\hat{d}^{+}_{2}
\ket{1,0,0}
=\frac{1}{8\pi\epsilon_{0}R^{3}}
\bra{1}
\hat{d}^{-}_{1}
\ket{0}
\bra{0}
\hat{d}^{+}_{2}
\ket{1},
\end{align}
In this way, all matrix elements can be reduced to obtaining a result for the dipole operator transition elements
\begin{align}
\bra{0}\hat{d}^{-}\ket{+}
&&
\bra{1}\hat{d}^{-}\ket{0}
&&
\bra{0}\hat{d}^{+}\ket{1}
&&
\bra{+}\hat{d}^{+}\ket{0}.
\end{align}
We can evaluate these matrix elements by rewriting the dipole operator
\begin{align}
\hat{d}^{+}=-\frac{1}{\sqrt{2}}\left(\hat{d}^{x}+\mathrm{i}\hat{d}^{y}\right)
=
-\frac{q}{\sqrt{2}}\left(\hat{x}+i\hat{y}\right)
=
-\vec{e}_{+}\cdot\hat{\vec{d}}
=
-q\vec{e}_{+}\cdot\hat{\vec{r}}
=
-q\vec{e}_{+}\cdot\hat{\vec{e}}_{r}\cdot r
\end{align}
where we have used $\vec{e}_{+}=\frac{1}{\sqrt{2}}\left(\vec{e}_{x}+i\vec{e}_{y}\right)$. Similar expression can be derived for $\hat{d}^{-}$. Additionally, we can decompose the position unit vector using the spherical harmonics $Y_{l,m}$ (see \cite{Foot2011})
\begin{align}
\vec{e}_{r}=\sqrt{\frac{4\pi}{3}}\left[
Y_{1,-1}\frac{\vec{e}_{x}+\mathrm{i}\vec{e}_{y}}{\sqrt{2}}
+
Y_{1,1}\frac{-\vec{e}_{x}+\mathrm{i}\vec{e}_{y}}{\sqrt{2}}
+
Y_{1,0}\vec{e}_{z}
\right].
\end{align}
To calculate the transition matrix element between two states $\ket{\Psi}$ and $\ket{\Psi^{\prime}}$ we will consider the angular and radial parts separately:
\begin{align}
\bra{\Psi^{\prime}}\hat{d}^{\pm}\ket{\Psi}
&=
-q\vec{e}_{\pm}\bra{\Psi^{\prime}}\hat{\vec{e}}_{r}\cdot r\ket{\Psi}
\\
&=
-q\bra{\psi^{\prime}}C_{1,\pm 1}\ket{\psi}
\int_{0}^{\infty}dr R^{\prime}_{n,l^{\prime}}(r)r R_{n,l}(r)
\end{align}
In the above equation, $\ket{\psi}$ and $\ket{\psi^{\prime}}$ denote the angular and $R^{\prime}_{n,l^{\prime}},  R_{n,l}$ the radial wavefunctions. The latter is equal to
\begin{align}
\xi\equiv\int_{0}^{\infty}dr R^{\prime}_{n,l^{\prime}}(r)r R_{n,l}(r)
\end{align}
and is constant over all transitions relevant here, so we simply set it to $\xi$.
For the angular part, we must evaluate the following expression
\begin{align}
\bra{\psi^{\prime}}\hat{d}^{\pm}\ket{\psi}
=
q\bra{\psi^{\prime}}C_{1,\pm 1}\ket{\psi}
\\
C_{l,m}=\sqrt{\frac{4\pi}{2l+1}}Y_{l,m}.
\end{align}
where $q=-e$ is the electron charge. Matrix elements of this kind can be evaluated using Wigner symbols (see \cite{Sobelman1979})
\begin{align}
\bra{\psi^{\prime}}C_{k,p}\ket{\psi}
=&
\delta_{S^{\prime},S}\sqrt{(2J^{\prime}+1)(2J+1)(2L^{\prime}+1)(2L+1)}(-1)^{M^{\prime}-S}
\\
&\times
\begin{pmatrix}
J^{\prime} & J & k\\
-M^{\prime} & M & p
\end{pmatrix}
\begin{pmatrix}
L^{\prime} & k & L\\
0 & 0 & 0
\end{pmatrix}
\begin{Bmatrix}
L^{\prime} & L & k\\
J & J^{\prime} & S
\end{Bmatrix}.
\end{align}
The brackets in the second line are referred to as Wigner-3j-Symbols and Wigner-6j-Symbols, respectively, and the variables contained in the above equation represent the quantum numbers of both states.
Using the atomic states in question, we arrive at (the common factor of $\xi$ is omitted)
\begin{align}
\bra{0}\hat{d}^{-}\ket{+}&=\frac{e}{\sqrt{3}}
&&
\bra{1}\hat{d}^{-}\ket{0}=-\frac{e}{3}
\\
\bra{0}\hat{d}^{+}\ket{1}&=\frac{e}{3}
&&
\bra{+}\hat{d}^{+}\ket{0}=-\frac{e}{\sqrt{3}}.
\end{align}
Now we can calculate explicitly the transition elements from above, starting from the unit of energy $J$:
\begin{align}
J&=-\bra{0,1,0}\hat{V}_{1,2}\ket{1,0,0}
=-\frac{1}{8\pi\epsilon_{0}R^{3}}
\bra{1}
\hat{d}^{-}_{1}
\ket{0}
\bra{0}
\hat{d}^{+}_{2}
\ket{1}
\\
&=\frac{e^{2}}{72\pi\epsilon_{0}R^{3}}
\end{align}
Analogously, we calculate the indirect transport processes from above, $h_{1\rightarrow 2\rightarrow 3}$
\begin{align}
h_{1\rightarrow 2\rightarrow 3}&=\bra{0,0,1}\hat{H}\ket{0,+,0}\bra{0,+,0}\hat{H}\ket{1,0,0}
\\
&=\bra{0,0,1}\hat{V}_{2,3}\ket{0,+,0}\bra{0,+,0}\hat{V}_{1,2}\ket{1,0,0}
\\
&=\left(\frac{1}{4\pi\epsilon_{0}R^{3}}\frac{3}{2}\right)^{2}
\bra{0,0,1}
\hat{d}^{-}_{2}\hat{d}^{-}_{3}\mathrm{e}^{2\iup\phi_{2,3}}
\ket{0,+,0}
\bra{0,+,0}
\hat{d}^{+}_{1}\hat{d}^{+}_{2}\mathrm{e}^{-2\iup\phi_{1,2}}
\ket{1,0,0}
\\
&=\left(\frac{1}{4\pi\epsilon_{0}R^{3}}\frac{3}{2}\right)^{2}\mathrm{e}^{-4\iup\alpha}\frac{e^{4}}{27}
\\
&=27J^{2}\mathrm{e}^{-4\iup\alpha}.
\end{align}
Similarly, one can evaluate the density-density terms $U_{i}$. 
\\
So far we have performed the derivation in the single-excitation manifold of the Hilbert space. In the case of two excitations on the triangle, all $h_{i\rightarrow j\rightarrow k}$ vanish, whereas $h_{i\rightarrow j\rightarrow i}$ remains non-zero if sites $i$ and $k$ are occupied. 
\\
Using this we arrive at the Hamiltonian in hard-core boson language as already shown in the main text
\begin{align}
\hat{H}&=-J\sum_{i\ne j=1}^{3}\bd_{i}\b_{j}+h.c.
-
2gJ\sum_{i\neq j\neq k}^{3}
\bd_{i}\b_{j}\left(1-\n_{k}\right)\mathrm{e}^{-4 i\epsilon_{i,j,k}\alpha}+h.c.
-
2gJ\sum_{i\neq k}\n_{i}\left(1-\n_{k}\right)
\end{align}
where we again introduce $g=\frac{27J}{2\Delta}$ as a dimensionless parameter.

\section{Local current vortices in the insulating regime}\label{appendix_vortices}

\begin{figure}[htb]
	\begin{center}
	\includegraphics[width=0.5\columnwidth]{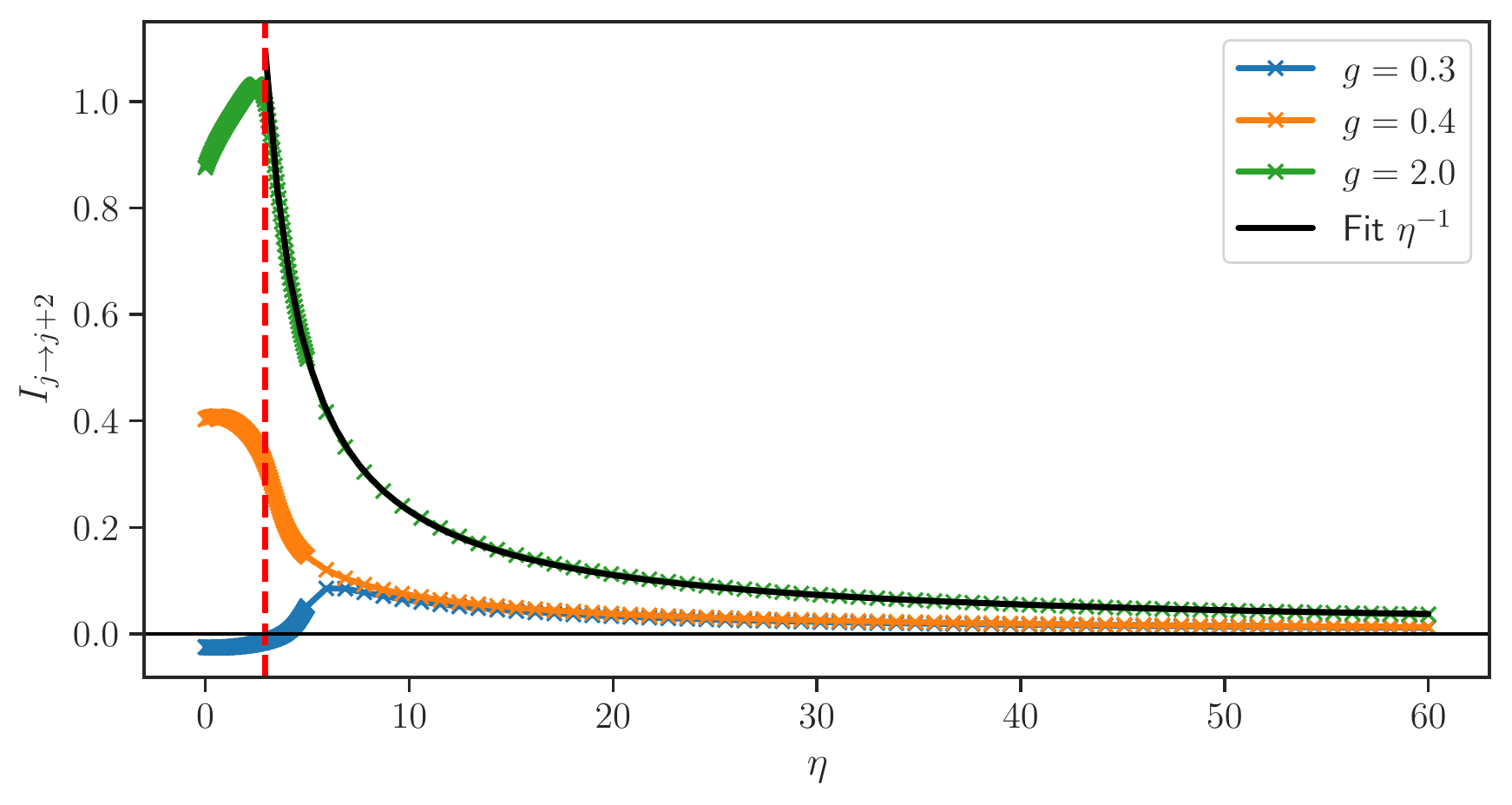}
	\end{center}
	\caption{Current $I_{j\rightarrow j+2}$ between nearest neighbor sites along the upper sub-chain for different values of $g$ as function of density-density repulsion strength $\eta$. One can clearly see the $1/\eta$ scaling as the system enters into the density-ordered phase. The red dotted line corresponds to the transition point for $g=2$ as marked in Fig.~\ref{fig:phase-diagramm} in the main text.}
	\label{fig:currents_insulating_regime}
\end{figure}

We have seen in Sect.\ref{Sect:density-driven-gauge} that in the density-ordered phase of large $\eta$ an oscillatory 
regular flux pattern of the quantum gauge field is induced with long-range correlations. A measurable consequence of this is the formation of a vortex lattice of local currents as seen in Fig.\ref{fig:emerging_gauge_field}b. Although there is no net current in an
insulating phase with density order and first-order nearest neighbor correlations are suppressed, their amplitudes decays only algebraically with 
the density-density repulsion  $\eta$. 
Thus local vortex currents can emerge even in the insulating phase. Their amplitude is expected to
decrease as $1/\eta$, which is verified by the very good agreement of the numerical data for the current $I_{j\to j+2}$ with an $1/r$ fit in Fig.~\ref{fig:currents_insulating_regime}. We marked the transition for $g= 2$ (green curve) by the red dotted line.
%

\bibliographystyle{iopart-num}
\bibliography{Rydberg-zigzag}


\end{document}